\documentclass[english,showpacs,preprintnumbers,nofootinbib,amsmath,amssymb]{revtex4}
\usepackage[T1]{fontenc}
\usepackage[latin9]{inputenc}
\usepackage{amsmath}
\usepackage{amssymb}
\usepackage{graphicx}

\makeatletter
\@ifundefined{textcolor}{}
{%
 \definecolor{BLACK}{gray}{0}
 \definecolor{WHITE}{gray}{1}
 \definecolor{RED}{rgb}{1,0,0}
 \definecolor{GREEN}{rgb}{0,1,0}
 \definecolor{BLUE}{rgb}{0,0,1}
 \definecolor{CYAN}{cmyk}{1,0,0,0}
 \definecolor{MAGENTA}{cmyk}{0,1,0,0}
 \definecolor{YELLOW}{cmyk}{0,0,1,0}
}

\usepackage{babel}

\@ifundefined{textcolor}{}{%
 \definecolor{BLACK}{gray}{0}
 \definecolor{WHITE}{gray}{1}
 \definecolor{RED}{rgb}{1,0,0}
 \definecolor{GREEN}{rgb}{0,1,0}
 \definecolor{BLUE}{rgb}{0,0,1}
 \definecolor{CYAN}{cmyk}{1,0,0,0}
 \definecolor{MAGENTA}{cmyk}{0,1,0,0}
 \definecolor{YELLOW}{cmyk}{0,0,1,0}
}

\@ifundefined{textcolor}{}{%
 \definecolor{BLACK}{gray}{0}
 \definecolor{WHITE}{gray}{1}
 \definecolor{RED}{rgb}{1,0,0}
 \definecolor{GREEN}{rgb}{0,1,0}
 \definecolor{BLUE}{rgb}{0,0,1}
 \definecolor{CYAN}{cmyk}{1,0,0,0}
 \definecolor{MAGENTA}{cmyk}{0,1,0,0}
 \definecolor{YELLOW}{cmyk}{0,0,1,0}
}

\@ifundefined{definecolor}{\usepackage{color}}{}
\usepackage{rotating}\usepackage{MnSymbol}

\newcommand{\beq}{\begin{equation}}
\newcommand{\eeq}{\end{equation}}
\newcommand{\bma}{\begin{math}}
\newcommand{\ema}{\end{math}}
\newcommand{\beqa}{\begin{eqnarray}}
\newcommand{\eeqa}{\end{eqnarray}}
\def\opone{\le\textbf{}\textbf{}avevmode\hbox{\small1\kern-3.8pt\normalsize1}}

\newcommand{\be}[1]{
{\marginpar{{\scriptsize\ \\ \ #1}}}
\begin{eqnarray} \mbox{$\label{#1}$} }

\newcommand{\ee}{\end{eqnarray}}

\newcommand{\ket}[1]{|#1 \rangle }

\newcommand{\id}{{\mathbf 1}}

\newcommand{\bl}{{\bf l}}

\allowdisplaybreaks[4]

\newlength{\cwidth}

\newlength{\xw}
\newcommand{\compp}[9]{%
\begin{matrix}
#4 & #2 & #1\\
#7 & #5 & #3\\
#9 & #8 & #6\\
\end{matrix}
}

\renewcommand{\compp}[9]{%
\begin{matrix}
\makebox[.7cm]{$#4$} & \makebox[.7cm]{$#2$} & \makebox[.7cm]{$#1$} \\
\makebox[.7cm]{$#7$} & \makebox[.7cm]{$#5$} & \makebox[.7cm]{$#3$} \\
\makebox[.7cm]{$#9$} & \makebox[.7cm]{$#8$} & \makebox[.7cm]{$#6$} \\
\end{matrix}
}

\newcommand{\nesw}{\neswarrow}
\newcommand{\nwse}{\nwsearrow}
\newcommand{\symd}{\neswarrow\hspace{-3.4mm}\nwsearrow\hspace{0mm}}

\renewcommand{\nesw}{%
\hspace{1 mm}
\raisebox{-1.3 mm}{\begin{rotate}{45} $\longleftrightarrow$ \end{rotate}}
\hspace{3.9 mm}}
\renewcommand{\nwse}{%
\hspace{4.9 mm}
\begin{rotate}{135} $\longleftrightarrow$ \end{rotate}}

\renewcommand{\symd}{%
\hspace{1 mm}
\raisebox{-1.3mm}{\begin{rotate}{45} $\longleftrightarrow$ \end{rotate}}
\hspace{3.9 mm}
\begin{rotate}{135} $\longleftrightarrow$ \end{rotate}
}

\allowdisplaybreaks[4]

\usepackage{babel}

\makeatother

\usepackage{babel}
\begin{document}

\title{Integrability in anyonic quantum spin chains via a composite height
model}

\author{Paata Kakashvili}
\altaffiliation[Current address: ]{Center for Materials Theory, Rutgers University, Piscataway,
New Jersey 08854, USA%
}
\affiliation{Nordita, Roslagstullsbacken 23, SE-106 91 Stockholm, Sweden}

\author{Eddy Ardonne}
\affiliation{Nordita, Roslagstullsbacken 23, SE-106 91 Stockholm, Sweden}

\date{\today}
\begin{abstract}
Recently, properties of collective states of interacting non-abelian
anyons have attracted a considerable attention. We study an extension
of the `golden chain model', where two- and three-body interactions
are competing. Upon fine-tuning the interaction, the model is integrable.
This provides an additional integrable point of the model, on top
of the integrable point, when the three-body interaction is absent.
To solve the model, we construct a new, integrable height model, in
the spirit of the restricted solid-on-solid model solved by Andrews,
Baxter and Forrester. The heights in our model live on both the sites
and links of the square lattice. The model is solved by means of the
corner transfer matrix method. We find a connection between local
height probabilities and characters of a conformal field theory governing
the critical properties at the integrable point. In the antiferromagnetic
regime, the criticality is described by the $Z_{k}$ parafermion conformal
field theory, while the $\frac{su(2)_{1}\times su(2)_{1}\times su(2)_{k-2}}{su(2)_{k}}$
coset conformal field theory describes the ferromagnetic regime. 
\end{abstract}

\pacs{05.30.Pr, 05.50.+q, 11.25.Hf, 02.30.Ik}

\maketitle

\section{Introduction}

The last half decade has seen a big increase in the interest of topological
phases of matter. In this paper, we will study a model which is inspired
by the prototype of a topological phase, namely the (fractional) quantum
Hall effect. It has been conjectured that there exist fractional quantum
Hall states with excitations which exhibit non-abelian statistics
\cite{mr91}. One of the key properties of this type of excitations,
called non-abelian anyons, is that a topological state with a number
of non-abelian anyons present, is degenerate. The number of degenerate
states is exponential, while the energy splitting in real systems
decays exponentially with the average distance between the non-abelian
anyons.

An important question which raises itself is what happens if the anyons
are close to one another, such that they start interacting. To this
end, a one-dimensional (1D) model of interacting anyons was constructed
in Ref. \cite{ftl07}, called the `golden chain', because it was based
on Fibonacci anyons. The philosophy behind this model was to stay
as closely as possible to a Heisenberg model of interacting spins.
In fact, the golden chain is precisely that, a Heisenberg model with
two-body nearest neighbor interactions, but for anyons instead of
spins. The phase diagram of such anyonic Heisenberg models turns out
to be rich, even richer than the phase diagrams of the ordinary spin
case. We will not embark on a long discussion of the phase diagrams
of these models here, but focus in the next section on one particular
example of interest for the current paper, an extension of the golden
chain model with competing two- and three-body interactions \cite{taf08}.
We would like to point out that studying the effects of interacting
anyons in 1D gives insight into the fate of interacting anyons in
two-dimensional (2D) systems. The interactions between the anyons
can nucleate a new topological liquid, and the collective behavior
of the 1D chain describes the boundary between the original and nucleated
topological phases, see Refs. \cite{gat09,lpt11} and Refs. \cite{gs09,bsh09}
for related work.

An interesting property of the anyonic chain models is that they exhibit
(fine-tuned) points, at which it is possible to solve the model exactly.
Obviously, having access to an exact solution, even though such a
solution is only available at special points, greatly enhances the
understanding of the model. The golden chain (with a two-body interaction),
and its cousins which are obtained by replacing the Fibonacci anyons
by anyons based on $su(2)_{k}$, can be mapped onto a two-dimensional
classical statistical mechanics model, namely the restricted solid-on-solid
(RSOS) model, introduced and solved by Andrews, Baxter and Forrester
(ABF) \cite{abf84}. This RSOS model consists of heights living on
the sites of the square lattice. Plaquettes are weighted depending
on the heights of the sites forming the plaquettes. To solve the model,
ABF employed the so-called corner transfer matrix (CTM) method \cite{book:b82}.
It was found that the model exhibits various ordered phases, separated
by critical points. The connection between the critical exponents
and conformal field theory (CFT) was made by Huse \cite{h84}. The
anyonic chains correspond to the RSOS model at the critical point,
and are therefor critical themselves, and governed by the same CFT
\cite{ftl07}.

The CTM method allows one to calculate, in the limit of infinite lattice
size, the probability for a site in the bulk to have a particular
height. Interestingly, it has been observed that off-critical local
height probabilities of integrable models are intimately related to
partition functions of the associated critical theories in a finite
box with appropriate boundary conditions \cite{sb89}. Moreover, close
to critical points, these height probabilities are given in terms
of characters of a CFT, which describes the critical behavior of the
model \cite{djk87}. The characters stemming from the height probabilities
in the RSOS model, are for instance given in Ref.~\cite{kkm93}.
These characters can be interpreted in terms of fractional exclusion
statistics \cite{h91}, or more specifically, a non-abelian version
thereof \cite{s97,gs99}. In light of the current paper, we would
like to point out that the height probabilities one obtains for a
finite system, correspond to finitized characters. Interestingly,
the opposite `ends' of these finitized characters correspond to the
full characters of two different conformal field theories, in the
large size limit. These two different conformal field theories describe
the critical properties of different critical points of the (RSOS)
model. In terms of the anyonic chains, these two critical points are
related to each other by changing the overall sign of the interaction.

In this paper, we will examine an integrable point of anyonic chains
with competing two- and three-body interaction terms. To solve the
model, we introduce a new statistical mechanics model, which builds
on the RSOS model, which has six different types of plaquettes. In
our new model, we will combine four plaquettes of the RSOS model,
and shift two of the plaquette weights, to obtain a non-trivial generalization.
Via this procedure, one obtains a model with sixty-six different types
of plaquettes. This composite height model gives rise to the anyonic
chain Hamiltonians of interest, via the usual `anisotropic limit'.
To solve the model, we will follow the work of ABF, and employ the
corner transfer matrix method to calculate the height probabilities.

A very closely related loop model has been studied in the literature
\cite{ijs09,ijs10}. In fact, the $R$-matrix, used to construct the
row-to-row transfer matrix, has the same underlying algebraic structure,
namely that of the Temperley-Lieb algebra, and the construction of
our 2D model is inspired by the work in Refs. \cite{ijs09,ijs10}.
However, the quantum chain of that work is defined on a completely
different Hilbert space, in comparison to the anyonic chains. In our
case, the Hilbert space does not have a tensor product decomposition,
which hinders solving the model by means of the Bethe Ansatz, which
is the method used in Refs. \cite{ijs09,ijs10} (in their representation,
the Hilbert space does have a tensor product decomposition). The Hilbert
spaces of the anyonic chains exhibit a non-local, topological symmetry,
giving rise to a topological quantum number, which can be used to
label the eigenstates of Hamiltonians respecting this symmetry (see,
for instance Refs. \cite{ftl07,longsu2k}). In gapped phases where
the ground state breaks this symmetry, one finds additional degeneracies.
This additional structure, which is intimately tied to the non-abelian
nature of the anyons, seems to be absent in the work of Refs.~\cite{ijs09,ijs10}.

The outline of the paper is as follows. In Section \ref{sec:anyon-chains},
we briefly introduce the anyonic chain Hamiltonians, starting with
the original golden chain, and its generalization by introducing the
three-body interaction. We also briefly discuss the anyonic chains
based on $su(2)_{k}$ anyons. In Section \ref{sec:statmech}, we discuss
the connection of the golden chain with the RSOS model. The following
Section \ref{sec:composite-model} contains the definition of our
new, composite height model, which builds on the RSOS model. The corner
transfer matrix method is described in Section \ref{sec:ctm}, which
leads to the expressions for the height probabilities. These are used
in the Section \ref{sec:phases}, to obtain information about the
various phases of the model. In Section \ref{sec:HeightProbabilities},
we calculate off-critical local height probabilities in different
phases and show that they are given in terms of characters of the
CFTs, which govern the corresponding critical points. We conclude
in Section \ref{sec:conclusions}. In Appendix \ref{app:plaquettes},
we explicitly give the different types of plaquettes of the composite
height model. Appendix \ref{app:ctm-properties} contains various
limits of the CTM's, which are used in the main text. Appendix \ref{app:p-one-limit}
deals with a certain limit of the plaquettes. Finally, in Appendix
\ref{app:cft-connection}, we give the details of the connection of
the various heigh probabilities with CFT characters.

\section{Anyon chains with competing interactions}

\label{sec:anyon-chains}

\subsection{The golden chain}

The first model of interacting anyons, introduced in Ref. \cite{ftl07},
was dubbed the `golden chain'. In this model, so-called Fibonacci
anyons interact in basically the same way as spins in the Heisenberg
spin chain, namely, an energy is assigned depending on the overall
spin state of two interacting particles. In the $S=1/2$ Heisenberg
chain, the $S=0$ state of two neighboring spins is favored for antiferromagnetic
interactions, while for ferromagnetic interactions, the $S=1$ channel
for neighboring spins is favored.

To explain the interaction of the golden chain, we will first briefly
introduce the notion of (Fibonacci) anyons. A rather extensive introduction
on this topic can be found in Ref. \cite{ttw08}. The Fibonacci anyon
model consists of two types of particles: the `trivial' or vacuum
particle $\id$, and the Fibonacci particle $\tau$. As with ordinary
spins, one can combine, or fuse these particles, and decompose the
product. This fusion product is the direct analog of taking tensor
products of spins. Contrary to $su(2)$ spin, there is only a finite
number of types of particles. In addition, there is no internal quantum
number, such as $s_{z}$. The reason for this will become clear shortly.

The rules for combining the anyons in the Fibonacci model are as follows
$\id\times\id=\id$, $\id\times\tau=\tau$ and $\tau\times\tau=\id+\tau$,
the latter being the only non-trivial fusion rule.

\begin{figure}[ht]
\includegraphics{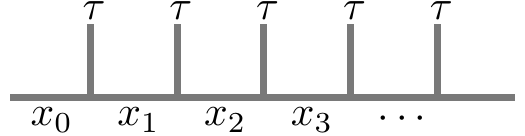} \caption{The fusion chain consisting of Fibonacci anyons. Consistent labelings
$(x_{0},x_{1},...,x_{L})$ form the Hilbert space of the anyonic chain.}

\label{fig:fusion-chain} 
\end{figure}

Let us take a chain of $L$ $\tau$ anyons. To describe the Hilbert
space of this system, it is easiest to think in terms of a so-called
fusion chain, as depicted in figure \ref{fig:fusion-chain}. 

This fusion chain consists of labelled lines, $L$ incoming lines
labelled $\tau$, which represent the Fibonacci anyons that form the
chain. The lines connecting these incoming $\tau$ anyons are labelled
$x_{0}$, $x_{1}$, \emph{etc}. These labels, which can take the values
$\id$ and $\tau$, are the `degrees of freedom'. The set of consistent
labelings $(x_{0},x_{1},\ldots,x_{L})$ forms the Hilbert space of
the chain. For a labeling to be consistent, the fusion rules have
to be satisfied at every vertex. This means that $x_{i+1}$ has to
be in the fusion of $x_{i}\times\tau$. This means that one can not
have $x_{i}=x_{i+1}=\id$, because this would violate the fusion rule
$\id\times\tau=\tau$. Apart from the constraint that no two neighboring
labels can both take the value $\id$, the labelings are arbitrary.
Because of the constraint, the size of the Hilbert space grows as
$d^{L}$, where $d<2$. It is not so hard to convince oneself that
in fact $d$ is the golden ratio, $d=\varphi=(1+\sqrt{5})/2$. In
the remainder of the description, we will assume periodic boundary
conditions, $x_{0}=x_{L}$. In this case, the size of the Hilbert
space is given by ${\rm dim}\mathcal{H}_{L}={\rm Fib}(L+1)+{\rm Fib}(L-1)$,
where ${\rm Fib}(n)$ is the $n^{{\rm th}}$ Fibonacci number, defined
by ${\rm Fib}(n)={\rm Fib}(n-1)+{\rm Fib}(n-2)$ and the initial conditions
${\rm Fib}(0)=0$ and ${\rm Fib}(1)=1$. Loosely speaking, one can
say that each Fibonacci anyon has a fractional number of degrees of
freedom, namely $d$, explaining the absence of an internal quantum
number. More importantly, one can not assign a local Hilbert space
to each anyon. This is the reason we had to resort to the fusion chain
to describe the Hilbert space, which can not be described as a tensor
product of local Hilbert spaces, as is the case for ordinary spin
chains.

We would like to point out that in the description of the Hilbert
space, we did not make use of the braid properties of the Fibonacci
anyons. Often, the braid properties are used to define the concept
of non-abelian statistics. What we have done here instead, is to use
the fusion properties of the Fibonacci anyons, which make non-abelian
statistics possible. In particular, in order for non-abelian statistics
to be possible, one needs a Hilbert space whose dimension is at least
two. This in turn is possible, if one considers particles (anyons)
which have multiple fusion channels upon fusion with another particle,
such as $\tau\times\tau=\id+\tau$. In the construction of interacting
anyonic chains, it is the presence of multiple fusion channels which
is the key property of non-abelian statistics which is utilized. One
can consider models in which the explicit, non-abelian braid properties
are used to define the Hamiltonian (see for instance \cite{ttw08}),
but that is not the route will take in the present paper.

We turn our attention to the description of the Hamiltonian. We first
concentrate on the Hamiltonian of the original golden chain model.
The interaction between two anyons depends on their overall fusion
channel. Favoring the overall fusion channel of two neighboring anyons
to be $\id$ will be called antiferromagnetic, while favoring the
$\tau$ channel will be called ferromagnetic interaction.

\begin{figure}[ht]
\includegraphics{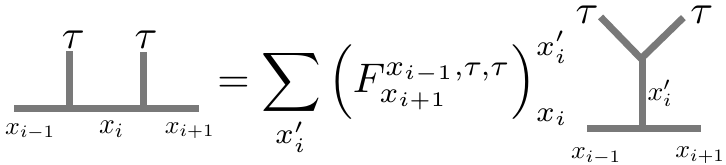} \caption{The $F$-matrix elements as a local basis transformation.}

\label{fig:f-matrix} 
\end{figure}

In our description of the Hilbert space in terms of a fusion chain,
the fusion channel of two neighboring anyons is not explicit, because
the lines associated with these two anyons do not meet in one vertex.
One can, however, perform a local basis transformation, which makes
this fusion channel explicit. The matrix describing this basis transformation
is called the $F$-matrix, which is the direct analog of the $6j$-symbols
in the case of $su(2)$ spin. These describe the change of basis between
the following two possible ways of describing the Hilbert space of
three spins: $(S_{1}\otimes S_{2})\otimes S_{3}$ and $S_{1}\otimes(S_{2}\otimes S_{3})$.
In figure \ref{fig:f-matrix}, we depict the $F$-matrix in terms
of the fusion-chain pictures. 

After performing this local basis transformation, the fusion channel
of the two neighboring anyons is explicit and it is given by $x'_{i}$.
Based on the value of this label (which can be $\id$ or $\tau$),
we can now assign the energy by projecting onto the $\id$ or $\tau$
channel. All that is left to do is to perform one more basis transformation,
to go back to the original basis.

To make the model completely explicit, we have to give the values
of the $F$-matrix elements. For the anyon models based on $su(2)$,
there exist explicit formulas for the $F$-matrices, see for instance
Ref.~\cite{kr88}. We will not go into the details here, but simply
give the results for the $F$-symbols we need. Note that there is
some degree of freedom in the $F$-symbols, the explicit form we give
fixes this. The $F$-symbols we need are specified by specifying the
values of $(x_{i-1},x_{i},x_{i+1})$, which can take the values $\{(\id,\tau,\id),(\id,\tau,\tau),(\tau,\tau,\id),(\tau,\id,\tau),(\tau,\tau,\tau)\}$.
In addition, we have to specify the degrees of freedom after the basis
transformation, $(x_{i-1},x'_{i},x_{i+1})$, which can take the values
$\{(\id,\id,\id);(\id,\tau,\tau);(\tau,\tau,\id);(\tau,\id,\tau),(\tau,\tau,\tau)\}$.
Using this ordering of the states, the $F$-matrix takes the form
(highlighting the block structure) 
\begin{equation}
F=\begin{pmatrix}1\\
 & 1\\
 &  & 1\\
 &  &  & d^{-1} & d^{-\frac{1}{2}}\\
 &  &  & d^{-\frac{1}{2}} & -d^{-1}
\end{pmatrix}\ .
\end{equation}
 The inverse transformation is given by the same matrix, because $F=F^{-1}$,
as is easily checked by using $d^{2}=1+d$. We can now easily form
the local projection operators $P_{{\rm 2-body}}^{(\tau)}$ and $P_{{\rm 2-body}}^{(\id)}$,
which project onto the $\tau$ and $\id$ channels, \emph{i.e.} give
energy to these channels. Hence, $P_{{\rm 2-body}}^{(\tau)}$ corresponds
to the antiferromagnetic interaction. Explicitly, these projection
matrices take the form $P_{{\rm 2-body}}^{(\tau)}=F\cdot{\rm diag}(0,1,1,0,1)\cdot F$
and $P_{{\rm 2-body}}^{(\id)}=F\cdot{\rm diag}(1,0,0,1,0)\cdot F$.
The components of these matrices read $\bigl(F_{x_{i+1}}^{x_{i-1},\tau,\tau}\bigr)_{\tau}^{\tilde{x}_{i}}\bigl(F_{x_{i+1}}^{x_{i-1},\tau,\tau}\bigr)_{x_{i}}^{\tau}$
and $\bigl(F_{x_{i+1}}^{x_{i-1},\tau,\tau}\bigr)_{\id}^{\tilde{x}_{i}}\bigl(F_{x_{i+1}}^{x_{i-1},\tau,\tau}\bigr)_{x_{i}}^{\id}$,
respectively. Explicitly written out, this becomes 
\begin{align}
P_{{\rm 2-body}}^{(\id)} & =\begin{pmatrix}1\\
 & 0\\
 &  & 0\\
 &  &  & d^{-2} & d^{-\frac{3}{2}}\\
 &  &  & d^{-\frac{3}{2}} & d^{-1}
\end{pmatrix} & P_{{\rm 2-body}}^{(\tau)} & =\begin{pmatrix}0\\
 & 1\\
 &  & 1\\
 &  &  & d^{-1} & -d^{-\frac{3}{2}}\\
 &  &  & -d^{-\frac{3}{2}} & d^{-2}
\end{pmatrix}\ .
\end{align}
 One can easily check that $P_{{\rm 2-body}}^{(\id)}+P_{{\rm 2-body}}^{(\tau)}=\openone$,
where $\openone$ is the identity matrix.

We can now write down the golden chain Hamiltonian as the sum of the
projection operators $P_{{\rm 2-body},i}^{(\tau)}$, $H=J_{2}\sum_{i=1}^{L}P_{{\rm 2-body},i}^{(\tau)}$,
where the projector $P_{{\rm 2-body},i}^{(\tau)}$ assigns a positive
energy if the anyons $i$ and $i+1$ are in the $\tau$ channel. In
the original golden chain paper \cite{ftl07}, it was shown numerically
that this model is critical for either sign of the interaction. Moreover,
the central charge was determined via the entanglement entropy, resulting
in $c=7/10$ and $c=4/5$ for antiferromagnetic $(J_{2}=1)$ and ferromagnetic
$(J_{2}=-1)$ interactions, respectively. Exact diagonalization of
the model showed that for antiferromagnetic interactions, the low-lying
part of the spectrum can be described in terms of the minimal model
$\mathcal{M}_{4,5}$, describing the tri-critical Ising model. The
critical model describing the ferromagnetic system is that of the
$Z_{3}$ parafermions.

It was subsequently realized that the golden chain Hamiltonian can
in fact be obtained from an exactly solvable model, a particular version
of the restricted solid-on-solid models \cite{abf84}. These models
exhibit various ordered phases, separated by critical points. It is
these critical points the golden chain can be mapped to. As a result,
one can obtain information about the critical theory of the golden
chain, by studying the critical behavior of the RSOS models. We will
discuss these RSOS models, and their connection to the anyonic chain
Hamiltonians in more detail in the next section.

\subsection{Competing interactions}

Having introduced the golden chain Hamiltonian, in which the anyons
interact via a two-body nearest-neighbor interaction, we now consider
the effect of introducing a three-body interaction, which was first
considered in Ref. \cite{taf08}.

It is well known that if one adds a three-body term (with large enough
coupling) to the $S=1/2$ Heisenberg antiferromagnet, a gap opens,
and one enters the Majumdar-Ghosh (MG) phase \cite{mg69a,mg69b}.
The phase diagram of the Heisenberg chain with competing nearest-neighbor
two- and three-body interactions is rather rich. Similarly, it was
expected that the phase diagram of the golden chain with competing
two and three-body interactions is rich as well. This model was studied
in Ref. \cite{taf08}, indeed finding an interesting phase diagram
(see Fig. \ref{fig:phasediagram}), which we briefly review below,
after introducing the details of the three-body interaction.

\begin{figure}
\includegraphics{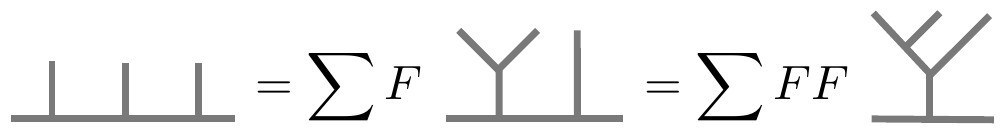} \caption{The $F$ transformations needed for the three-body interaction.}

\label{fig:3body} 
\end{figure}

In order to find the fusion channel of three neighboring anyons, we
first have to perform two $F$ transformations, after which this fusion
channel is explicit. One can then project onto the desired channel,
and go back to the original basis. This was explained in detail in
Ref. \cite{ttw08}. The schematics of the basis transformation is
given in figure \ref{fig:3body}. 

The three-body interaction will depend on four labels $(x_{i-1},x_{i},x_{i+1},x_{i+2})$.
To give the interaction matrix, we use the following basis $\{(\id,\tau,\tau,\id);(\id,\tau,\id,\tau),(\id,\tau,\tau,\tau);(\tau,\id,\tau,\id),(\tau,\tau,\tau,\id);(\tau,\id,\tau,\tau),(\tau,\tau,\id,\tau),(\tau,\tau,\tau,\tau)\}$.
In this basis, the three-body projectors take the following form 
\begin{align}
P_{{\rm 3-body}}^{(\id)} & =\begin{pmatrix}1\\
 & 0 & 0\\
 & 0 & 0\\
 &  &  & 0 & 0\\
 &  &  & 0 & 0\\
 &  &  &  &  & d^{-2} & d^{-2} & -d^{-\frac{5}{2}}\\
 &  &  &  &  & d^{-2} & d^{-2} & -d^{-\frac{5}{2}}\\
 &  &  &  &  & -d^{-\frac{5}{2}} & -d^{-\frac{5}{2}} & d^{-3}
\end{pmatrix} & P_{{\rm 3-body}}^{(\tau)} & =\begin{pmatrix}0\\
 & 1 & 0\\
 & 0 & 1\\
 &  &  & 1 & 0\\
 &  &  & 0 & 1\\
 &  &  &  &  & d^{-1} & -d^{-2} & d^{-\frac{5}{2}}\\
 &  &  &  &  & -d^{-2} & d^{-1} & d^{-\frac{5}{2}}\\
 &  &  &  &  & d^{-\frac{5}{2}} & d^{-\frac{5}{2}} & 2d^{-2}
\end{pmatrix}\ .
\end{align}
 In terms of these projectors, the most general interaction we can
write down takes the form 
\begin{equation}
H_{J_{2},J_{3}}=\sum_{i=1}^{L}\cos\theta P_{{\rm 2-body},i}^{(\tau)}+\sin\theta P_{{\rm 3-body},i}^{(\tau)}\ .
\end{equation}
 In this equation, the three-body projectors act on the quadruples
$(x_{i-1},x_{i},x_{i+1},x_{i+2})$, while the two-body projectors
act on triples $(x_{i-1},x_{i},x_{i+1})$. Moreover, we assume periodic
boundary conditions, $x_{i+L}=x_{i}$, and we introduced the couplings
$J_{2}=\cos\theta$ and $J_{3}=\sin\theta$.

\subsection{Phase diagram of the $J_{2}-J_{3}$ model}

\begin{figure}[ht]
\includegraphics[width=10cm]{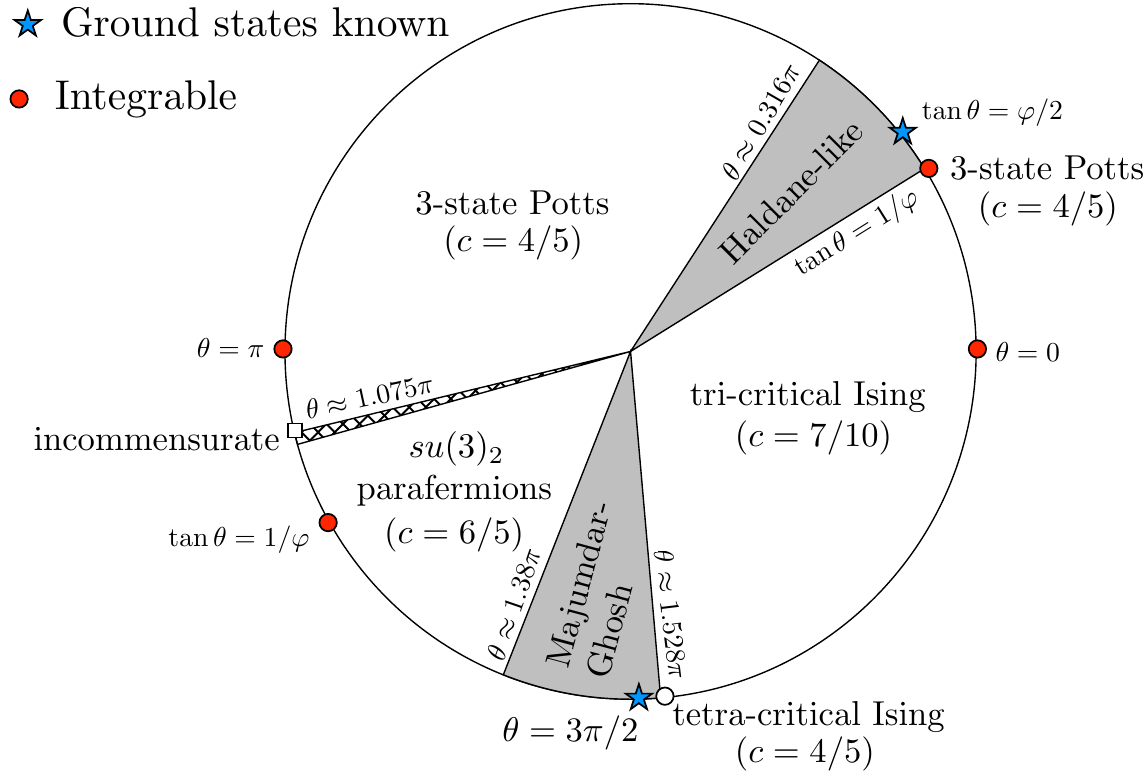} \caption{Phase diagram of the $J_{2}-J_{3}$ model; $J_{2}=\cos\theta$, $J_{3}=\sin\theta$.}

\label{fig:phasediagram} 
\end{figure}

We now briefly describe the phase diagram of the $J_{2}-J_{3}$ model,
as a function of the angle $\theta$. For more details, we refer to
Refs. \cite{taf08,longj3}. The phase diagram of this model is shown
in figure \ref{fig:phasediagram}. 

The angle $\theta=0$ corresponds to the original, two-body golden
chain model, which is critical and described by the tri-critical Ising
model. For a finite range of both positive and negative angles, this
behavior persists, so we have an extended range of critical behavior.
At the angle given by $\tan\theta=1/\varphi$, with both the two-
and three-body interactions antiferromagnetic, there is a phase transition
to a gapped phase. In this gapped phase, the ground state is two-fold
degenerate, with all the ground states occurring at zero momentum.
It turns out that at the transition point, which is described by the
$Z_{3}$ parafermion theory, a non-local symmetry, dubbed topological
symmetry, is broken.

The critical phase around $\theta=0$ also gives way for a gapped
phase if $\theta$ is decreased, namely around $\theta\approx-0.472\pi$.
In the resulting gapped phase, both the spatial and topological symmetries
are broken, giving rise to four (dimerized) ground states. This phase
is the anyonic equivalent of the Majumdar-Ghosh phase in spin-1/2
chains with a purely three-body interaction. The phase transition
from the tri-critical Ising region to the MG-like phase is described
by the tetra-critical Ising model.

The critical behavior at $\theta=\pi$, the original golden chain
with ferromagnetic interactions is described in terms of the $Z_{3}$
parafermions (the critical behavior of the 3-state Potts model). This
point is part of an extended critical region, which extends all the
way to the gapped phase in the region when both couplings are antiferromagnetic.
This endpoint of that gapped phase is around $\theta\approx0.316\pi$.

The other end point of the extended critical region containing the
point $\theta=\pi$ is marked by a first order transition located
at $\theta\approx1.075\pi$ to a sliver of an incommensurate region,
which quickly gives way to an extended critical region. This region
has low-lying states at momenta $K=0,\pi/2,\pi,3\pi/2$, and was therefor
dubbed the $Z_{4}$ phase. This phase has a transition to the MG phase
around $\theta\approx1.38\pi$.

We will close this quick walk through the phase diagram by noting
that there are two special points which lie in the gapped phases,
namely at $\tan\theta=\varphi/2$ and $\theta=3\pi/2$. At these special
points, the ground states of the gapped phases are exactly degenerate
(as opposed to exponentially degenerate with system size), and moreover,
one can obtain these ground states explicitly. For more details about
the phase diagram and its peculiarities, we refer to Refs.\cite{taf08,longj3}.

In this paper, we will mainly concentrate on a special angle, given
by $\tan\theta=1/\varphi$, which corresponds to (on the one hand),
the transition between the extended critical region at antiferromagnetic
two-body interactions and the gapped phase obtained by introducing
the antiferromagnetic three-body interaction. Upon changing the sign
of the Hamiltonian, one ends up in the extended $Z_{4}$ critical
region. We will show in the next subsection, that by making use of
the integrable structure at $\theta=0$, one can show that the $J_{2}-J_{3}$
model is also integrable at $\tan\theta=1/\varphi$. We will confirm
that the critical point at $\tan\theta=1/\varphi$ (with both couplings
positive) is indeed described by the $Z_{3}$ parafermions. In addition,
we will show that for both couplings negative, the critical theory
describing the model is that of the Gepner parafermions related to
$su(3)_{2}$ \cite{g87}. This latter integrable point lies in an
extended critical region. Because there are no relevant operators
in the same symmetry sector as the ground state, this whole critical
region will be described by the same critical theory as the one we
found at the integrable point.

\subsection{Anyonic chains of $su(2)_{k}$ anyons}

In this subsection, we will describe the generalization of the golden
chain, where the Fibonacci anyons are replaced by more general types
of anyons. These more general anyons are of the type which is dubbed
$su(2)_{k}$, where $k$ is a positive integer. For arbitrary $k$,
this anyon theory has $k+1$ types of anyons, which can be labelled
in terms of an `angular momentum' $l$, which takes the values $l=0,\frac{1}{2},\ldots,\frac{k}{2}$.
For our present purposes, we are mainly interested in the fusion rules
of these anyons, and the associated $F$-symbols, which are necessary
to construct the Hamiltonians of interacting anyons of this type.

The fusion rules of the $su(2)_{k}$ anyons are derived from the tensor
products of spin representations of $SU(2)$. These have to be modified,
to take into account that in the anyon model, there is a highest angular
momentum. This generalization reads as follows. The fusion of two
anyons of type $j_{1}$ and $j_{2}$ is 
\begin{equation}
j_{1}\times j_{2}=\sum_{j_{3}=|j_{1}-j_{2}|}^{\min(j_{1}+j_{2},k-j_{1}-j_{2})}j_{3}\ ,\label{eq:fusionrules}
\end{equation}
 where the sum is either over the integers or half-integers. The only
difference between the tensor product rules for $SU(2)$ spins is
the upper bound. In particular, the case of the Fibonacci anyons corresponds
to $k=3$. In general, this theory has four anyons, $l=0,1/2,1,3/2$,
but because $k=3$ is odd, one can restrict oneself to the integer
subset (see Ref. \cite{ttw08} for details). This integer subset was
written as $\{\id,\tau\}$ in the previous subsections.

The model one now considers is the model where the constituent anyons
are the $l=\frac{1}{2}$ anyons of $su(2)_{k}$. The Hilbert space
consists of all labelings of the fusion tree in figure \ref{fig:fusion-chain},
but with the $\tau$ particles replaced by the $l=\frac{1}{2}$ anyons,
and at the vertices, the fusion rules in Eq. \eqref{eq:fusionrules}
have to be satisfied. The construction of the interaction matrices,
both for the two-body as well as three-body interactions, is identical
to the construction in the Fibonacci case. The only thing which has
to be changed is the $F$-matrix elements. Symbolically, we can write
the elements of the resulting projection matrices in the same way,
\begin{align}
\bigl(P_{{\rm 2-body},i}^{(1)}\bigr)_{x_{i-1},x_{i},x_{i+1}}^{x_{i-1},\tilde{x}_{i},x_{i+1}} & =\bigl(F_{x_{i+1}}^{x_{i-1},\frac{1}{2},\frac{1}{2}}\bigr)_{1}^{\tilde{x}_{i}}\bigl(F_{x_{i+1}}^{x_{i-1},\frac{1}{2},\frac{1}{2}}\bigr)_{x_{i}}^{1}\ ,\\
\bigl(P_{{\rm 3-body},i}^{(1/2)}\bigr)_{x_{i-1},x_{i},x_{i+1},x_{i+2}}^{x_{i-1},\tilde{x}_{i},\tilde{x}_{i+1},x_{i+2}} & =\sum_{x'_{i},x''_{i}}\;\bigl(F_{\tilde{x}_{i+1}}^{x_{i-1},\frac{1}{2},\frac{1}{2}}\bigr)_{x''_{i}}^{\tilde{x}_{i}}\bigl(F_{x_{i+2}}^{x_{i-1},x''_{i},\frac{1}{2}}\bigr)_{1/2}^{\tilde{x}_{i+1}}\bigl(F_{x_{i+2}}^{x_{i-1},x'_{i},\frac{1}{2}}\bigr)_{x_{i+1}}^{1/2}\bigl(F_{x_{i+1}}^{x_{i-1},\frac{1}{2},\frac{1}{2}}\bigr)_{x_{i}}^{x'_{i}}\ .
\end{align}
 Here, we wrote the projectors onto the spin-1 and spin-1/2 channels
for the two- and three-body interactions respectively. The $F$-matrices
themselves can be obtained from the explicit expressions, which were
derived in Ref. \cite{kr88}. The Hamiltonian for general $k$ now
reads 
\begin{equation}
H_{J_{2}-J_{3}}=\sum_{i}\cos\theta P_{{\rm 2-body},i}^{(1)}+\sin\theta P_{{\rm 3-body},i}^{(1/2)}\ ,
\end{equation}
 where the $k$ dependence is hidden in the detailed form of the projectors,
and of course in the Hilbert space itself.

The phase diagram of the general $k$ model has the same structure
as the phase diagram for $k=3$ in Fig.~\ref{fig:phasediagram}.
The extended critical region around $\theta=0$ is described by the
minimal model $\mathcal{M}_{k+1,k+2}$, the $k$-critical Ising model.
The critical phase around $\theta=\pi$ is described by the $Z_{k}$
parafermions (we refer to Ref. \cite{zf85} for a description of this
CFT). At angles $\theta=0,\pi$, this follows from the integrability
of the RSOS model \cite{ftl07}.

We introduce the notation $d_{k}=2\cos\bigl(\pi/(k+2)\bigr)$ for
the quantum dimension of the spin$-1/2$ anyon of the $su(2)_{k}$
anyon model. Below, we will show that the angles given by $\tan\theta=(d_{k}^{2}-1)/d_{k}^{2}$
are special, because we can obtain the critical behavior by mapping
the model to a new integrable generalization of the RSOS model. For
the resulting $\theta$ in the range $0\leq\theta\leq\pi/2$, this
integrable point is the transition from the extended critical region
to a gapped phase. From numerics, it was already obtained that this
critical behavior is described by the $Z_{k}$ parafermion theory
\cite{taf08}. Below, we show that this indeed follows by exactly
solving the model. The opposite point lies within the so-called $Z_{4}$
critical region, with low-lying states at the momenta $K=0,\pi/2,\pi,3\pi/2$.
For this critical region, the numerical results were less clear, but
our analysis of the integrable 2D classical statistical mechanics
model, which we introduce in this paper, shows that the conformal
field theory description of this phase is in terms of a diagonal coset
model, namely $\frac{su(2)_{1}\times su(2)_{1}\times su(2)_{k-2}}{su(2)_{k}}$.
For $k=3$, this model reduces to the $su(3)_{2}$ parafermion CFT.
Bordering this extended critical phase is the analog of the Majumdar-Ghosh
phase around $\theta=3\pi/2$, which also borders the extended critical
phase around $\theta=0$. The phase transition between the latter
two is described by the $k+1$-critical Ising model.

\section{Connection with $2D$ statistical mechanics models}

\label{sec:statmech}

\subsection{Integrability of the Golden chain model}

In the original paper \cite{ftl07}, it was pointed out that the golden
chain Hamiltonian can be solved exactly by mapping it onto the restricted
solid-on-solid model, which was exactly solved by Andrews, Baxter
and Forrester \cite{abf84} by means of the corner transfer matrix
method \cite{book:b82}. In particular, the two-body terms $P_{{\rm 2-body},i}^{(1)}$
in the Hamiltonian at $\theta=0$, $H_{J_{2}=1,J_{3}=0}=\sum_{i=1}^{L}P_{{\rm 2-body},i}^{(1)}$,
can be related to generators of the Temperley-Lieb algebra $e(i)$,
namely $e(i)=d_{k}\bigl(\openone-P_{{\rm 2-body},i}^{(1)}\bigr)$.

The Temperley-Lieb algebra generators satisfy the relations

\begin{eqnarray}
e^{2}(i) & = & d_{k}\ e(i)\nonumber \\
e(i)e(i\pm1)e(i) & = & e(i)\label{eq:TLAlgebra}\\
{}[e(i),e(j)] & = & 0\quad{\rm for}\ |i-j|\geq2\ .\nonumber 
\end{eqnarray}

\begin{figure}
\includegraphics{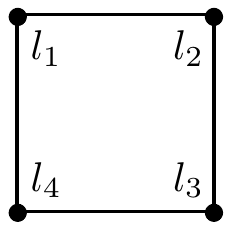} \caption{The plaquette weight $W(l_{1},l_{2},l_{3},l_{4})$, which in the anisotropic
limit ($u\rightarrow0$) gives rise to the two-body Hamiltonian $H_{J_{2}=1,J_{3}=0}$.}
\label{fig:GC-plaquette} 
\end{figure}

The action of the Temperley-Lieb generators on the local degrees of
freedom labeling the states in the Hilbert space can be written, following
Pasquier \cite{p87}, as 
\begin{align}
e(i)\ket{x_{i-1},x_{i},x_{i+1}} & =\sum_{x'_{i}}\bigl(e(i)_{x_{i-1}}^{x_{i+1}}\bigr)_{x_{i}}^{x'_{i}}\ket{x_{i-1},x'_{i},x_{i+1}}\\
\bigl(e(i)_{x_{i-1}}^{x_{i+1}}\bigr)_{x_{i}}^{x'_{i}} & =\delta_{x_{i-1},x{}_{i+1}}\sqrt{\frac{S_{0,x_{i}}S_{0,x'_{i}}}{S_{0,x_{i-1}}S_{0,x_{i+1}}}}\ ,\label{eq:anyonrep}
\end{align}
 where $S_{i,j}$ are the elements of the modular $S$-matrix of the
$su(2)_{k}$ conformal field theory, which are labeled by $i,j=0,1/2,\ldots,k/2$,
corresponding to the different type of anyons in $su(2)_{k}$ theory.
Explicitly, one has 
\begin{equation}
S_{i,j}=\sqrt{\frac{2}{k+2}}\sin\Bigl(\frac{(2i+1)(2j+1)\pi}{k+2}\Bigr).
\end{equation}
 We can write down plaquette weights, or $R$-matrix elements of the
corresponding 2D classical statistical mechanics model, in terms of
the $e(i)$ and the identity operator $\openone$ as follows 
\begin{align}
R_{i}(u)_{\vec{x}}^{\vec{x}'} & =\Bigl(\frac{\sin(\frac{\pi}{k+2}-u)}{\sin(\frac{\pi}{k+2})}\openone_{\vec{x}}^{\vec{x}'}+\frac{\sin(u)}{\sin(\frac{\pi}{k+2})}e(i)_{\vec{x}}^{\vec{x}'}\Bigr)\equiv\Bigl(\prod_{j\neq i}\delta_{x'_{j},x_{j}}\Bigr)W(x_{i-1},x_{i}',x_{i+1},x_{i})\label{eq:ham-plaquettes}\\
e(i)_{\vec{x}}^{\vec{x}'} & =\Bigl(\prod_{j\neq i}\delta_{x'_{j},x_{j}}\Bigr)\bigl(e(i)_{x_{i-1}}^{x_{i+1}}\bigr)_{x_{i}}^{x'_{i}}\nonumber \\
\openone_{\vec{x}}^{\vec{x}'} & =\prod_{j}\delta_{x'_{j},x_{j}}\ .\nonumber 
\end{align}
 The subscript $i$ on the $R$-matrix labels the plaquette on which
it acts, while the argument $u$ is the so-called fugacity, and $W(x_{i-1},x_{i}',x_{i+1},x_{i})$
gives the weight of an elementary plaquette shown in Fig. \ref{fig:GC-plaquette}.
The above $R$-matrix can be shown to satisfy the Yang-Baxter equation:

\begin{equation}
R_{j}(u)R_{j+1}(u+v)R_{j}(v)=R_{j+1}(v)R_{j}(u+v)R_{j+1}(u)\ ,\label{eq:YB-equation}
\end{equation}
 by making use of the Temperley-Lieb algebra relations for the $e(i)$
in Eq. (\ref{eq:TLAlgebra}).

\begin{figure}[b]
\includegraphics[height=3cm]{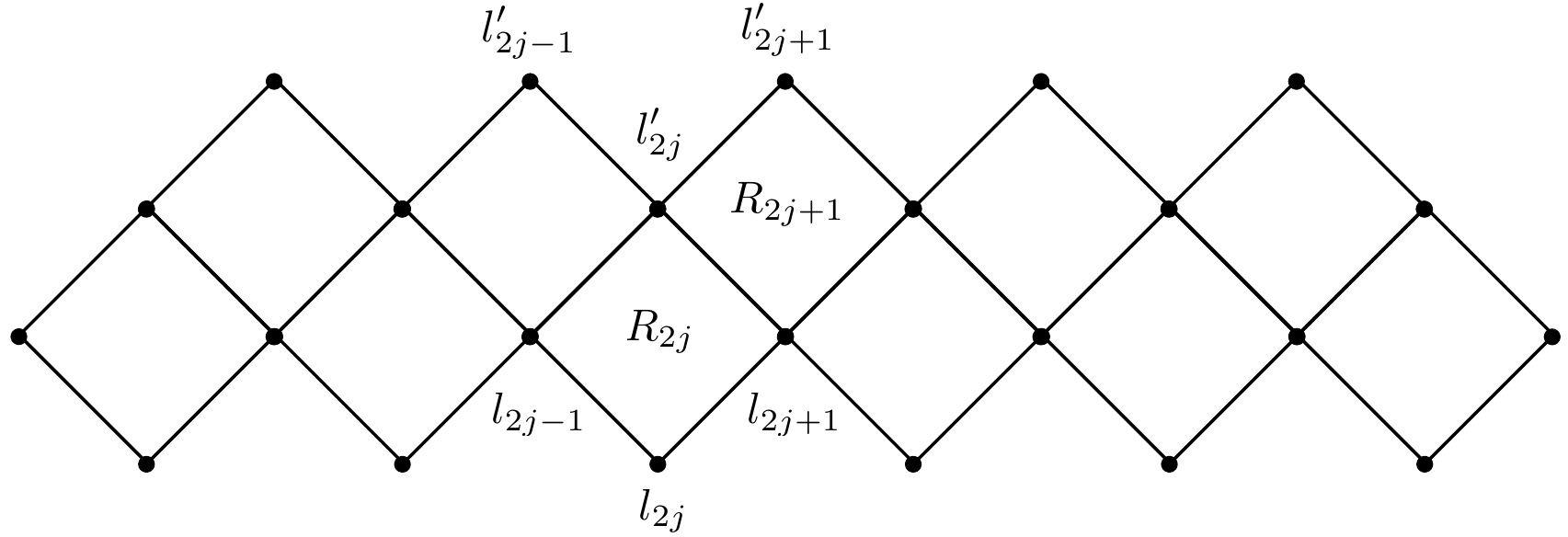}
\caption{`Two-row' transfer matrix, which plays a role of the discrete time-evolution
operator for the corresponding 1D quantum system. The transfer matrix
acts on the Hilbert space spanned by vectors $(...,l_{2j-1},l_{2j},l_{2j+1,}...)$.
(Imaginary) time runs vertically.}
\label{fig:TransferMatrix} 
\end{figure}

From the $R$-matrix, one can construct the `two-row' transfer matrix
(see Fig. \ref{fig:TransferMatrix}) 
\begin{equation}
T(u)=\prod_{j}R_{2j+1}(u)\prod_{j}R_{2j}(u)\ ,
\end{equation}
 which has a role of a discrete time-evolution operator for the corresponding
1D quantum system. The time-evolution operator acts on a Hilbert space,
which is spanned by vectors $(...,l_{2j-1},l_{2j},l_{2j+1,}...)$.
For the Golden chain model, this Hilbert space coincides with that
given in Fig. \ref{fig:fusion-chain}. We note that in making the connection between
the 2D statistical mechanics model and the 1D quantum Hamiltonian, we have
rotated the plaquettes by 45 degrees, or the time runs from the south-west
to north-east corner of the plaquette. In addition, we consider a `two-row' transfer
matrix, in order that in one discrete time step, all heights can evolve. This allows for
the possibility to obtain a translationally invariant Hamiltonian. The following calculation
shows that this is indeed the case.

If $R_{i}(u)$ satisfies the
Yang-Baxter equation, it follows that the $T$'s at different parameters
$u$ commute, allowing one to construct a Hamiltonian which can be
solved exactly. In particular, one writes $T(u)=e^{-uH+o(u^{2})}$,
which gives rise to the Hamiltonian 
\begin{equation}
H=-\frac{d\ln T(u)}{du}\bigr|_{u=0}=-\sum_{i}\frac{1}{R_{i}(u=0)}\frac{dR_{i}(u)}{du}\bigr|_{u=0}\ .\label{eq:anisotropic-limit}
\end{equation}
 Applying this construction to the $R$-matrix defined above, one
obtains 
\begin{equation}
H=\frac{2}{\tan(\pi/(k+2))}\sum_{i}\left(\frac{1}{2}-\frac{1}{d_{k}}e(i)\right)=\frac{2}{\tan(\pi/(k+2))}\sum_{i}\left(P_{{\rm 2-body},i}^{(1)}-\frac{1}{2}\right)\ ,\label{eq:AnisotropicLimit}
\end{equation}
 which is, up to a \emph{positive} scale factor and an overall shift,
equal to the golden chain Hamiltonian, $H_{J_{2}=1,J_{3}=0}$.

We now focus our attention on the RSOS model we referred to at the
beginning of this section. The model consists of height variables
(simply called `heights') located at the vertices of the square lattice.
The heights can take the values $l=1,2,\ldots,r-1$, where $r$ is
an arbitrary integer. We already noted that the heights correspond
to the different type of anyons, $0,1/2,\ldots,k/2$, where $k=r-2$.
The connection with the anyon Hamiltonian becomes complete by
following identification: $r\equiv k+2,$ and $l_{i}\equiv2x_{i}+1$,
where $l_{i}$ is the value of the height at the vertex $i$.

The heights have to satisfy the constraint that they differ by one
if they are nearest neighbors. Weights are assigned to the different
types of plaquettes, which we introduce below.

This model can be solved for a two-parameter family of weights, namely
for the parameter $u$, and an additional parameter $p$. This parameter
$p$ is the parameter which drives a phase transition, located at
$p=0$. The golden chain is related to the RSOS model at this critical
point.

\begin{figure}
\includegraphics{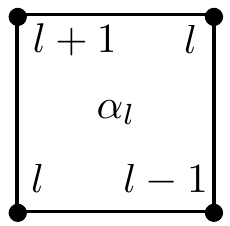} \hspace{0.5cm} \includegraphics{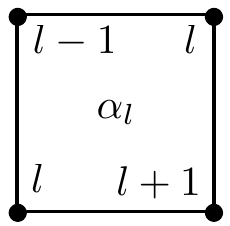}
\hspace{0.5cm} \includegraphics{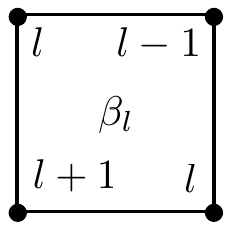} \hspace{0.5cm} \includegraphics{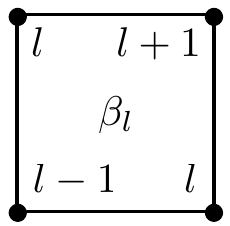}
\hspace{0.5cm} \includegraphics{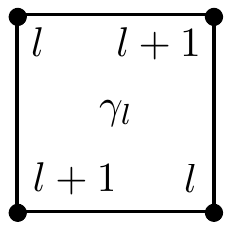} \hspace{0.5cm} \includegraphics{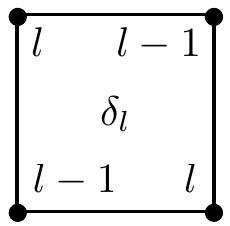}
\caption{Six height configurations occurring in the RSOS model.}

\label{fig:rsos-weights} 
\end{figure}

We now briefly describe the weights of the RSOS model in terms of
the parameters%
\footnote{The parameter $u$ is related to $v$ in Ref. \cite{abf84} by $u=\eta-v$.%
} $\eta$ (which is related to $r$, see below), $u$ and $p$. There
are six different arrangements of heights around a plaquette, as shown
in Fig. \ref{fig:rsos-weights}. The corresponding weights are given
in terms of elliptic functions as follows 
\begin{align}
\alpha_{l}(u) & =\frac{h(2\eta-u)}{h(2\eta)} & \beta_{l}(u) & =\frac{h(u)}{h(2\eta)}\frac{[h(w_{l-1})h(w_{l+1})]^{1/2}}{h(w_{l})}\label{eq:rsos-weights}\\
\gamma_{l}(u) & =\frac{h(w_{l}+u)}{h(w_{l})} & \delta_{l}(u) & =\frac{h(w_{l}-u)}{h(w_{l})}\ ,\nonumber 
\end{align}
 where $w_{l}=2\eta l$. The function $h(u)$ is given in terms of
elliptic theta functions, with argument $u$ and modulus $\tilde{k}=m^{2}$,
namely $h(u)=H(u)\Theta(u)$. The functions $H(u)$ and $\Theta(u)$
can be expressed in terms of the theta functions $\theta_{1}$ and
$\theta_{4}$, in particular, $H(u)=\theta_{1}(\frac{u\pi}{2K(m)},p)$
and $\Theta(u)=\theta_{4}(\frac{u\pi}{2K(m)},p)$, where $K(m)$ is
the complete elliptic integral of the first kind. The parameter $\eta$
is given in terms of $K$ and $r$, namely $\eta=K/r$. Furthermore,
$p$ can be expressed in terms of $m$ as $p=\exp\left[-\pi K'(m)/K(m)\right]$,
where $K'(m)=K(1-m)$. Using the product expansions of the elliptic
theta functions, one can write 
\begin{equation}
h(u)=2p^{1/4}\sin\left(\frac{\pi u}{2K(m)}\right)\prod_{n=1}^{\infty}(1-2p^{n}\cos\left(\frac{\pi u}{K(m)}\right)+p^{2n})(1-p^{2n})^{2}\ .
\end{equation}
 Note that we suppressed the dependence of the weights in Eq.~\eqref{eq:rsos-weights}
on $p$. We refer to chapter 15 of the book \cite{book:b82}, where
the properties of the elliptic functions used in this paper are analyzed.
It was shown in Ref.~\cite{abf84} that the weights given in Eq.~\eqref{eq:rsos-weights}
satisfy the Yang-Baxter equation for all $u$ and $p$.

Two phase transitions (in different regimes for $u$) occur for $p=0$,
which implies that $m=0$. These two critical points correspond to
the integrable point of the golden chain model with $\tan\theta=0$
(\emph{i.e.} the angles $\theta=0,\pi$ of the $J_{2}-J_{3}$ model).
Thus, to relate the weights in Eq.~\eqref{eq:rsos-weights} at the
critical point to the plaquette weights obtained from the golden chain
model, we take the limit $\lim_{p\rightarrow0}h(u_{1})/h(u_{2})=\sin(u_{1})/\sin(u_{2})$,
where we used that $K(0)=\pi/2$.

Using this limit, one finds, at the critical point, that 
\begin{align}
\alpha_{l} & =\frac{\sin(\frac{\pi}{k+2}-u)}{\sin(\frac{\pi}{k+2})} & \beta_{l} & =\frac{\sin(u)}{\sin(\frac{\pi}{k+2})}\frac{[\sin(\frac{(l-1)\pi}{k+2})\sin(\frac{(l+1)\pi}{k+2})]^{1/2}}{\sin(\frac{l\pi}{k+2})}\label{eq:anyon-weights}\\
\gamma_{l} & =\frac{\sin(\frac{l\pi}{k+2}+u)}{\sin(\frac{l\pi}{k+2})} & \delta_{l} & =\frac{\sin(\frac{l\pi}{k+2}-u)}{\sin(\frac{l\pi}{k+2})}\ ,\nonumber 
\end{align}
 which are identical to those obtained from Eq. (\ref{eq:ham-plaquettes}).

The last statement can be verified by considering the explicit form
of the $R$-matrix in Eq. \eqref{eq:ham-plaquettes}. The $\alpha$
type plaquettes are obtained from the first term in Eq. \eqref{eq:ham-plaquettes},
when $x_{i-1}\neq x_{i+1}$. The $\beta$ plaquettes are the `off-diagonal'
terms, with $x_{i}\neq x'_{i}$, which can occur if $x_{i-1}=x_{i+1}$.
Only the second term in Eq. \eqref{eq:ham-plaquettes} contributes
to these plaquettes. Finally, the plaquettes of type $\gamma$ and
$\delta$ (which are diagonal, but also have $x_{i-1}=x_{i+1}$) receive
contributions from both terms in Eq. \eqref{eq:ham-plaquettes}.

Details of the connection between the critical behavior of the golden
chain can be found in the original paper \cite{ftl07}. For details
of the various phases of the RSOS model, we refer to Refs. \cite{abf84}
and \cite{h84}. Approaching the critical point $p=0$ from the positive
$p$ side, one finds that the observables of the model, such as the
height probabilities (see below), are given in terms of the tri-critical
Ising model for $u>0$, and the three-state Potts model for $u<0$.
This establishes the observed critical behavior of the golden chain
model \cite{ftl07}. In general, \emph{i.e.} for arbitrary $k$, this
generalizes to the $k$-critical Ising model for antiferromagnetic
interactions ($u>0$), and the $Z_{k}$ parafermions for ferromagnetic
($u<0$) interactions.

\section{Construction of a new composite height model}

\label{sec:composite-model}

\subsection{R-matrix for the composite model}

To identify a new integrable model, which corresponds to different
parameter regimes of the generalized golden chain model ($J_{2}-J_{3}$
model), we will use the ideas put forward in two papers by Ikhlef
\textit{et al.} \cite{ijs09,ijs10}. In these papers, a model closely
related to the anyonic chains is studied. Namely, the underlying algebraic
structure, the Temperley-Lieb algebra, is the same, but a different
representation is chosen. We consider the `anyonic representation'
(see Eq. \eqref{eq:anyonrep}), while Refs. \cite{ijs09,ijs10} consider
a spin-1/2-type representation. An important consequence of this difference
in representation is that we were forced to use the corner transfer
matrix method solve the model, as opposed to the Bethe Ansatz method
\cite{book:kbi93}, which was used in Refs. \cite{ijs09,ijs10}.

Below, we will introduce a new height model, which at its critical
point reduces to the anyonic chain at the integrable point. However,
we will also study the ordered phases of our new height model.

It was put forward in Ref. \cite{ijs09}, that one can use the $R$-matrix
in Eq. \eqref{eq:ham-plaquettes} to construct a composite $R$-matrix
$\tilde{R}$, which also satisfies the Yang-Baxter equation%
\footnote{This way of constructing a composite $R$-matrix reminds of the techniques
to construct higher spin, or `fused' models, see Refs. \cite{djk87,ks93}%
}. The composite $R$-matrix one has to consider takes the following
form 
\begin{equation}
\tilde{R}_{j}(u,\phi)=R_{2j+1}(u-\phi)R_{2j}(u)R_{2j+2}(u)R_{2j+1}(u+\phi)\ ,\label{eq:comp-R-matrix}
\end{equation}
 where we introduced an additional parameter $\phi$, ranging over
$0\leq\phi\leq\pi/2$. It can easily be shown that the $R$-matrix
\eqref{eq:comp-R-matrix} satisfies the Yang-Baxter equation. The
only ingredient needed to show this is that the original $R$-matrix
satisfies the Yang-Baxter equation itself. In what follows, we focus
on the case $\phi=\pi/2$, but we stress that the $R$-matrix in Eq.
\eqref{eq:comp-R-matrix} satisfies the Yang-Baxter equation for all
values of $\phi$. At $\phi=\pi/2$, the composite $R$-matrix leads
to particularly interesting points of the $J_{2}-J_{3}$ Hamiltonian,
namely the critical point between the `Haldane gap' phase and the
extended AFM critical region. The opposite point (considered in Refs.
\cite{ijs09,ijs10} in the different representation) lies within the
$Z_{4}$ critical region.

We now describe in some detail how to obtain the Hamiltonian of the
$J_{2}-J_{3}$ model, by taking the `anisotropic limit'. We start
by expanding the composite matrix $\tilde{R}_{i}(u,\phi)$ explicitly
(using the notation $\gamma=\pi/(k+2)$ and $e_{i}=e(i)$), 
\begin{equation}
\begin{split}\sin^{4}(\gamma)\tilde{R}_{i}(u,\phi)= & -\bigl(\sin(\gamma-u)^{2}\sin(\phi+\gamma-u)\sin(\phi-\gamma+u)\bigr)\openone\\
 & -\bigl(\sin(u)\sin(\gamma-u)\sin(\phi+\gamma-u)\sin(\phi-\gamma+u)\bigr)(e_{2i}+e_{2i+2})\\
 & +\frac{1}{2}\sin(u)\sin(\gamma-u)\bigl(1+2\cos(2\phi)-\cos(2\gamma)-\cos(2(\gamma-u))-\cos(2u)\bigr)e_{2i+1}\\
 & +\bigl(\sin(u)\sin(\gamma-u)\sin(\phi-u)\sin(\phi-\gamma+u)\bigr)(e_{2i}e_{2i+1}+e_{2i+2}e_{2i+1})\\
 & +\bigl(\sin(u)\sin(\gamma-u)\sin(\phi+u)\sin(\phi+\gamma-u)\bigr)(e_{2i+1}e_{2i}+e_{2i+1}e_{2i+2})\\
 & -\sin(u)^{2}\sin(\phi+\gamma-u)\sin(\phi-\gamma+u)e_{2i}e_{2i+2}\\
 & +\sin(u)^{2}\sin(\phi-u)\sin(\phi-\gamma+u)e_{2i}e_{2i+2}e_{2i+1}\\
 & +\sin(u)^{2}\sin(\phi+u)\sin(\phi+\gamma-u)e_{2i+1}e_{2i}e_{2i+2}\\
 & -\sin(u)^{2}\sin(\phi+u)\sin(\phi-u)e_{2i+1}e_{2i}e_{2i+2}e_{2i+1}
\end{split}
\ .
\end{equation}
 We note that the coefficients of the terms $e_{2i}e_{2i+1}e_{2i+2}$
and $e_{2i+2}e_{2i+1}e_{2i}$ are zero.

As we explained in the previous section, one can construct a Hamiltonian
related to this $R$-matrix via the `two-row' transfer matrix
$T(u)=\prod_{j}R_{2j+1}(u)\prod_{j}R_{2j}(u)=e^{-uH+o(u^{2})}$,
by taking the anisotropic limit, Eq. \eqref{eq:anisotropic-limit}.
Applying this procedure to the composite $R$-matrix $\tilde{R}_{i}(u,\phi)$,
one sees that in the `two-row' transfer matrix, see Fig.~\ref{fig:TransferMatrix},
one has to change $R$ to $\tilde{R}$. Consequently, it is not obvious that the
procedure yields a translationally invariant Hamiltonian. Applying the procedure,
one obtains 
\begin{equation}
\begin{split}H= & \sum_{i}\bigl(\frac{2\cos(2\gamma)-\cos(2\phi)-1}{2\sin(\gamma+\phi)\sin(\gamma-\phi)}\bigr)\openone-\frac{1}{\sin(\gamma)}(e_{i}+e_{i+1})\\
 & +\frac{\sin(\phi)}{\sin(\gamma)\sin(\phi+\gamma)\sin(\phi-\gamma)}\bigl(\sin(\phi)\cos(\gamma)(e_{i}e_{i+1}+e_{i+1}e_{i})+(-1)^{i}\cos(\phi)\sin(\gamma)(e_{i}e_{i+1}-e_{i+1}e_{i})\bigr)
\end{split}
\end{equation}
To make the connection with the anyonic chain, we 
focus on the case $\phi=\pi/2$, which gives (after dropping the irrelevant constant) 
\begin{equation}
H=\frac{2}{\sin(2\gamma)}\sum_{i}-\cos(\gamma)(e_{i}+e_{i+1})+(e_{i}e_{i+1}+e_{i}e_{i+1})\ .\label{eq:ham-int}
\end{equation}

We note that for $\phi=0$, we obtain the original golden chain model
$H=-\sum_{i}e_{i}$. In the case that $0<\phi<\pi/2$, the term $(e_{i}e_{i+1}-e_{i+1}e_{i})$
has a non-zero coefficient, which gives rise to a non-hermitian Hamiltonian,
and breaks the translational invariance.

To relate the Hamiltonian we just obtained from the composite $R$-matrix
to that of the $J_{2}-J_{3}$ model, we write the two- and three-body
projectors appearing in the $J_{2}-J_{3}$ anyonic Hamiltonian in
terms of the Temperely-Lieb generators $e_{i}$. In general, the projectors of
$p$ `spin-1/2' particles onto the `spin-$p$/2' channel, can be written
in terms of the $e_{i}$. Explicitly, for two
and three particles, one has (see also \cite{j83,w87})
\begin{align}
P_{{\rm 2-body},i}^{(1)} & =\openone-\frac{1}{d_{k}}e_{i}\ ,\nonumber \\
P_{{\rm 3-body},i}^{(3/2)} & =\openone-\frac{d_{k}}{d_{k}^{2}-1}(e_{i}+e_{i+1})+\frac{1}{d_{k}^{2}-1}(e_{i}e_{i+1}+e_{i+1}e_{i})\ ,\label{eq:jw-proj}
\end{align}
 where we remind the reader that $d_{k}=2\cos(\pi/(k+2))$. In the
anyonic spin chain the three-body interaction was written in terms
of the projector onto the spin-1/2 channel, which reads 
\begin{equation}
P_{{\rm 3-body},i}^{(1/2)}=\frac{d_{k}}{d_{k}^{2}-1}(e_{i}+e_{i+1})-\frac{1}{d_{k}^{2}-1}(e_{i}e_{i+1}+e_{i+1}e_{i})\ .
\end{equation}
 We can now write the $J_{2}-J_{3}$ model in two different ways,
namely 
\begin{align}
H_{J_{2}-J_{3}} & =\sum_{i}\cos(\theta)P_{{\rm 2-body},i}^{(1)}+\sin(\theta)P_{{\rm 3-body},i}^{(1/2)}\ ,\\
H_{e} & =\sum_{i}\cos(\theta_{e})(e_{i}+e_{i+1})+\sin(\theta_{e})(e_{i}e_{i+1}+e_{i+1}e_{i})\ .
\end{align}
By making use of the projectors in Eq. \eqref{eq:jw-proj}, we find
the following relation between the angles $\theta$ and $\theta_{e}$
\begin{align}
\tan\theta & =\frac{(d_{k}^{2}-1)\tan\theta_{e}}{2d_{k}(1+d_{k}\tan\theta_{e})} & \tan\theta_{e} & =\frac{(d_{k}^{2}-1)-2d_{k}^{2}\tan\theta}{2d_{k}\tan\theta}\ .\label{eq:theta-rel}
\end{align}
 The angle $\theta_{e}$ for which the model is integrable, $\theta_{e,{\rm int}}$,
can be read off from Eq. \eqref{eq:ham-int}, and is given by $\tan\theta_{e,{\rm int}}=-1/\cos\gamma=-2/d_{k}$.
This corresponds to the angle $\theta_{{\rm int}}$ in the $J_{2}-J_{3}$
model 
\begin{equation}
\tan\theta_{{\rm int}}=\frac{d_{k}^{2}-1}{d_{k}^{2}}
\end{equation}
 In particular, in the case $u>0$, we find that the corresponding
angle $\theta$ is given by $\theta=\arctan((d_{k}^{2}-1)/d_{k}^{2})+\pi$,
\emph{i.e.} when both the two- and three-body interactions are ferromagnetic.
When $u<0$, we have $\theta=\arctan((d_{k}^{2}-1)/d_{k}^{2})$, and
both interactions are antiferromagnetic.

\subsection{Constructing the composite height model}

As we described in the previous section for the original golden chain
model, we have to consider a more general two-dimensional height model,
in order to obtain the critical behavior of the $J_{2}-J_{3}$ model
at the integrable points. The plaquettes of this new integrable height
model, which we introduce below, reduce to the composite $R$-matrix
described above at the critical point $p=0$.

\begin{figure}[ht]
\includegraphics{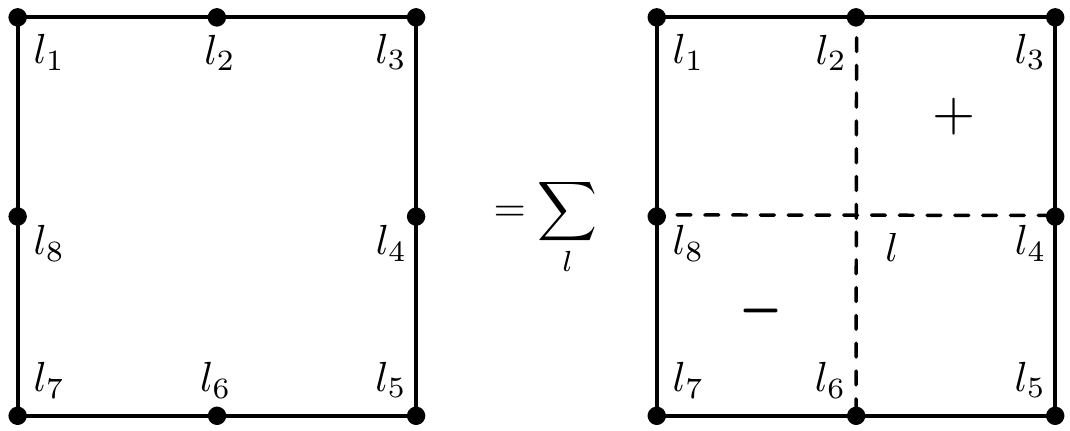} \caption{The plaquette weight $\tilde{W}(l_{1},l_{2},l_{3},l_{4},l_{5},l_{6},l_{7},l_{8})$
for the composite model in terms of the weights of the RSOS model.}

\label{fig:Gen-plaquette} 
\end{figure}

From the construction of the $R$-matrix, we know how to construct
the plaquette-weights for the composite 2D classical statistical mechanics
model from those of the RSOS model described in the previous section.
The new plaquettes consist of four plaquettes in the original RSOS
model, as depicted in figure \ref{fig:Gen-plaquette}. The model again
lives on the square lattice, where both the vertices as well as the
middle of the links have a height variable. Two neighboring heights
have to differ by one, as in the original model. One can think of
these plaquettes as composite plaquettes, whose weights depend on
the four original plaquettes forming the composite one. The original
plaquettes each contribute to the weight of the composite plaquette,
but two of the plaquettes have an appropriate `shift', as in the construction
of the composite $R$-matrix (see Eq. (\ref{eq:comp-R-matrix})).

Using the six different types of plaquettes of the RSOS model, it
turns out that one can construct 66 different types of composite plaquettes,
which are given in Appendix \ref{sec:plaquettes}. Not all the weights
of these plaquettes are different. Moreover, they satisfy certain
symmetries, in the same way as the original RSOS model. The actual
number of different plaquettes depends on the parameter $r$, which
determines the number of allowed heights, via $l_{i}=1,2,\ldots,r-1$.

The composite weights $\tilde{W}$ (see Fig. \eqref{fig:Gen-plaquette})
in term of the RSOS weights have the following form: 
\begin{eqnarray}
 &  & \tilde{W}(l_{2j-1},l_{2j}',l_{2j+1}',l_{2j+2}',l_{2j+3},l_{2j+2},l_{2j+1},l_{2j})=\\*
 &  & \sum_{l}W^{-}(l_{2j},l,l_{2j+2},l_{2j+1})W(l_{2j-1},l_{2j}',l,l_{2j})W(l,l_{2j+2}',l_{2j+3},l_{2j+2})W^{+}(l_{2j}',l_{2j+1}',l_{2j+2}',l),\nonumber 
\end{eqnarray}
 where $W^{\pm}(u)=W(u\pm K)$ are the weights of the shifted plaquettes
(see Fig. \ref{fig:Gen-plaquette}). We note that each of the plaquettes
forming the composite plaquette is of type $\alpha_{l},\beta_{l},\gamma_{l},\delta_{l}$,
with the appropriate values of $l$, namely 
\begin{align}
W(l+1,l,l-1,l) & =W(l-1,l,l+1,l)=\alpha_{l}(u)\ ,\nonumber \\
W(l,l-1,l,l+1) & =W(l+1,l,l-1,l)=\beta_{l}(u)\ ,\\
W(l,l+1,l,l+1) & =\gamma_{l}(u)\ ,\nonumber \\
W(l,l-1,l,l-1) & =\delta_{l}(u)\ ,\nonumber 
\end{align}
 where the explicit expressions of these weights, in terms of the
parameters $u$ and $p$, are given in Eq. (\ref{eq:rsos-weights}).

Taking into account the quasi-periodic properties of $h(u)$, we have
chosen the shift $\phi$ to equal $K$. Again, similar to the RSOS
model, we will be interested in $p\rightarrow0$ (critical) limit
and note that in this limit, $\phi=K=\pi/2$. Due to the symmetry
properties of elliptic functions we only need to consider the region
$2\eta-K<u<2\eta+K$, which naturally breaks into two domains according
to the sign of $u$ (cf. Ref. \cite{abf84}):

\begin{eqnarray}
{\cal D}_{1} & : & 0<u<2\eta+K=(2+r)\eta\ ,\\
{\cal D}_{2} & : & 2\eta-K=(2-r)\eta<u<0\ .
\end{eqnarray}
 The different signs of the fugacity correspond to ferro- (${\cal D}_{1}$)
and antiferromagnetic (${\cal D}_{2}$) regimes, which in the anisotropic
limit give rise to the integrable point of the generalized $J_{2}-J_{3}$
model, for both signs of the interaction.

We have now completely specified our new height model. We employ the
corner transfer matrix method, which is described in the next section,
to solve it. The main interest is to calculate local height probabilities
in different domains for general $p$, \emph{e.i.} away from criticality.
It has been observed that these off-critical ($p\ne0$) local height
probabilities of an integrable lattice model can be mapped to partition
functions of the corresponding critical theory ($p=0$) in a finite
box with appropriate boundary conditions \cite{sb89}. This mapping
is realized if one properly relates $p$, which plays the role of
temperature, to the finite size $L$ of the critical system. In addition,
it has been realized that the local critical probabilities can be
written using characters of the underling CFT \cite{djk87}. Relying
on these observations, we identify CFTs describing critical theories
of the generalized anyon model ($p=0$) in subsequent sections. In
particular, we calculate the off-critical local height probabilities
and relate them to characters of a CFT, which governs the critical
properties of the generalized anyon model (as well as the generalized
RSOS model at $p=0$).

\section{Corner transfer matrix method and local height probabilities}

\label{sec:ctm}

\subsection{Definition of corner transfer matrices}

To exactly solve the generalized model, we use the corner transfer
matrix method in analogy to the solution of the RSOS model by Andrews,
Baxter and Forrester \cite{abf84}. Here we give a short account of
the method and turn the interested reader to literature for more details
\cite{book:b82,b07}. The object of interest is the local height probability
$P_{a}$, which is the probability for a site to have height $a$.
This height probability is given by 
\begin{equation}
P_{a}=\frac{1}{Z}\sum_{\text{configurations}}(S_{a}\prod_{\text{\text{plaquettes}}}\tilde{W}(l_{j_{1}},l_{j_{2}},l_{j_{3}},l_{j_{4}},l_{j_{5}},l_{j_{6}},l_{j_{7}},l_{j_{8}}))\ ,
\end{equation}
 where product is over all plaquettes (faces) of the lattice and sum
runs over all allowed 2D height configurations, and 
\begin{equation}
(S_{a})_{\mathbf{l},\mathbf{l}'}=\delta(l_{1},a)\prod_{i=1}^{m}\delta(l_{i},l_{i}')\ .
\end{equation}
 The size of the system is parametrized by $m$, which should not
be confused with the $m$ related to the modulus $m^{2}=\tilde{k}$
of the theta-functions which appeared in section \ref{sec:statmech}.
The meaning of the indices $\bl,\bl'$ of the matrix $S$ will become
clear shortly.

\begin{figure}
\includegraphics[scale=0.7]{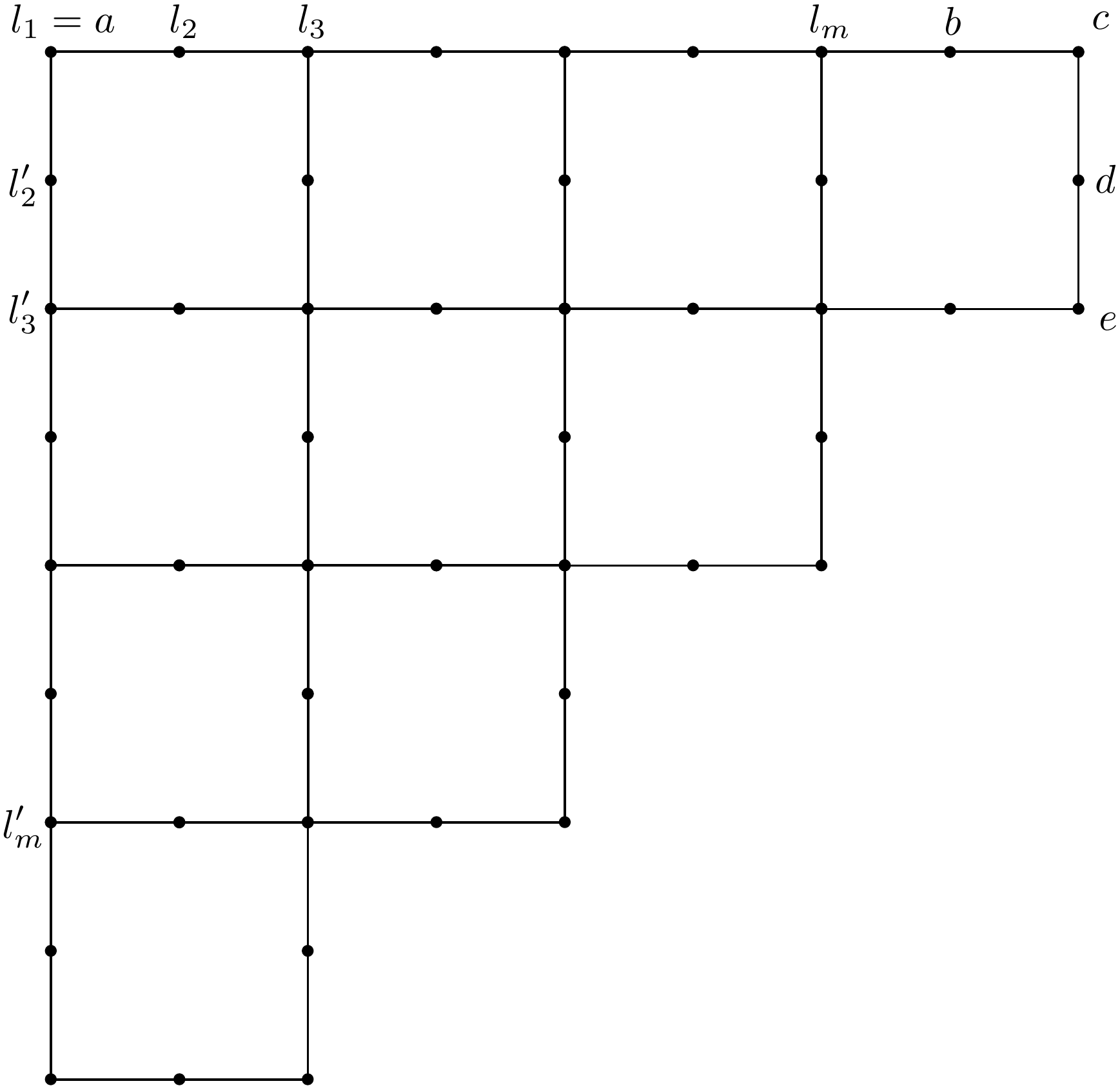} \caption{Corner transfer matrix $A$ for finite $m$, constructed using the
composite plaquettes. Heights $b,c,d,e$ are the boundary heights
fixed by a ground state pattern chosen.}

\label{fig:CTM} 
\end{figure}

The partition function $Z$, which is given by 
\begin{equation}
Z=\sum_{\text{configurations}}\prod_{\text{\text{plaquettes}}}\tilde{W}(l_{j_{1}},l_{j_{2}},l_{j_{3}},l_{j_{4}},l_{j_{5}},l_{j_{6}},l_{j_{7}},l_{j_{8}})\ ,\label{eq:PartitionFunction}
\end{equation}
 can be expressed as

\begin{equation}
Z=Tr(ABCD)\ ,
\end{equation}
 by introducing corner transfer matrices $A,B,C,D$, corresponding
to lower-right, upper-right, upper-left and lower-left quadrants of
the lattice (see Fig. \eqref{fig:CTM}, and text below for the precise
definition of $A,B,C,D$). These corner transfer matrices are analogs
of the row-to-row transfer matrix $T$, but instead of adding a row
to the lattice, they add a whole corner.

Finally, the local height probability can be written as

\begin{equation}
P_{a}=Tr(S_{a}ABCD)/Tr(ABCD)\ .\label{eq:LocalProb}
\end{equation}

We now have to introduce the notion of \emph{ground states}, which
will be used extensively in the subsequent analysis. Ground states
are those configurations of heights, which maximize the summand (or
minimize the ``energy'') in Eq. (\ref{eq:PartitionFunction}). These
ground states depend on $u$ in a way that in different domains of
$u$ (${\cal D}_{1}$ and ${\cal D}_{2}$) different ground state
types exist and they change discontinuously across the boundary between
different domains ($u=0$). These different ground states determine
different critical behavior in corresponding domains as we shall show
below.

The corner transfer matrices can be expressed using \emph{local} plaquette
(face) transfer matrices

\begin{eqnarray}
(U_{j})_{\mathbf{l},\mathbf{l}'} & = & \tilde{W}(l_{2j-1},l_{2j}',l_{2j+1}',l_{2j+2}',l_{2j+3},l_{2j+2},l_{2j+1},l_{2j})\prod_{i=1\ne2j,2j+1,2j+2}^{m}\delta(l_{i},l_{i}')\ ,\label{eq:defU}\\
(V_{j})_{\mathbf{l},\mathbf{l}'} & = & \tilde{W}(l_{2j+1}',l_{2j+2}',l_{2j+3},l_{2j+2},l_{2j+1},l_{2j},l_{2j-1},l_{2j}')\prod_{i=1\ne2j,2j+1,2j+2}^{m}\delta(l_{i},l_{i}')\ .\label{eq:defV}
\end{eqnarray}
 Here $j=1,...,(m+1)/2$, with $m$ odd, and $l_{m+1},l_{m+1}',l_{m+2},l_{m+2}',l_{m+3},l_{m+3}',l_{m+4}$
are boundary heights, which should be fixed to proper ground state
values corresponding to a considered domain of $u$. The matrix $U_{j}$
($V_{j}$) adds a plaquette to the lattice in the NE-SW (NW-SE) direction.

Using the definitions in Eqs. (\ref{eq:defU},\ref{eq:defV}), the
corner transfer matrix $A$ can be expressed as

\begin{equation}
A=F_{1}F_{2}...F_{(m+1)/2}\ ,\label{eq:cmtA}
\end{equation}
 where

\begin{equation}
F_{j}=U_{(m+1)/2}U_{(m-1)/2}...U_{j}\ .
\end{equation}

The corner transfer matrix $A$ (see Fig.~\ref{fig:CTM}) has rows
and columns, which are labeled by the values of the boundary heights,
collected in the vectors $\bl=(l_{1},l_{2},\ldots,l_{m})$ and $\bl'=(l'_{1},l'_{2},\ldots,l'_{m})$,
where $l'_{1}=l_{1}=a$, because these heights correspond to the same,
central height. In addition, we will use boundary conditions, such
that the heights on the four outermost diagonals are fixed to be $b,c,d,e$,
respectively.

The corner transfer matrices $B,C,D$ are expressed similarly to $A$
by replacing $U_{j}$ with $V_{j},U_{j}^{T}$and $V_{j}^{T}$, respectively.
In general, properties of the corner transfer matrices will depend
on symmetries of weights as well as those of ground states. In what
follows, we will be interested in infinite lattice limit, $m\rightarrow\infty$.

\subsection{Corner transfer matrices as exponentials.}

From the Yang-Baxter equation for weights (see Eq. (\ref{eq:YB-equation}))
follows a very important property of corner transfer matrices. In
the limit where the lattice size goes to infinity we can write (symbolically)
that 
\begin{equation}
\lim_{m\rightarrow\infty}B\, C=\lim_{n\rightarrow\infty}T^{n},
\end{equation}
 where $T$ is the row-to-row transfer matrix and $n$ is the number
of rows, covering a half plane in the limit $n\rightarrow\infty$.

We should note that the above relation is not valid for finite $m$
and $n$, since different boundary conditions are used to calculate
left- and right-hand sides of the equation (in fact, even the shapes
of the lattices differ). However, in the large $m,n$ limit this difference
becomes negligible. The Yang-Baxter equation ensures that the row-to-row
transfer matrices with different fugacities commute and, hence, the
product $B(u)C(v)$ depends only on the difference $u-v$ (modulus
the overall multiplicative factor). Similar equations that involve
other corner transfer matrices can be obtained by rotating the lattice
in steps of $\pi/2$. Using these properties one can show that the
corner transfer matrices have the following form (dropping irrelevant
multiplicative factors) \cite{book:b82}:

\begin{eqnarray}
A(u) & = & Q_{1}M_{1}e^{-u{\cal H}}Q_{2}^{-1}\ ,\label{eq:CTMExp}\\
B(u) & = & Q_{2}M_{2}e^{u{\cal H}}Q_{3}^{-1}\ ,\nonumber \\
C(u) & = & Q_{3}M_{3}e^{-u{\cal H}}Q_{4}^{-1}\ ,\nonumber \\
D(u) & = & Q_{4}M_{4}e^{u{\cal H}}Q_{1}^{-1}\ ,\nonumber 
\end{eqnarray}
 where the matrices ${\cal H},Q_{1},\cdots,Q_{4},M_{1},\cdots,M_{4}$
are independent of $u$ and can be chosen to commute with $S_{1},\cdots,S_{r-1}$.
In addition, the matrices ${\cal H},M_{1},\cdots,M_{4}$ are diagonal.

Using the identity (see Appendix \ref{cmtprop}) 
\begin{equation}
A(0)=Q_{1}M_{1}Q_{2}^{-1}=\openone\ ,
\end{equation}
 we immediately see that 
\begin{equation}
A(u)=Q_{2}e^{-u{\cal H}}Q_{2}^{-1}\ ,
\end{equation}
 which implies that the diagonal form of the corner transfer matrix
$A$ can be written as an exponential. From Eqs. (\ref{eq:LocalProb},\ref{eq:CTMExp})
we immediately see that

\begin{equation}
P_{a}=Tr(S_{a}M_{1}M_{2}M_{3}M_{4})/Tr(M_{1}M_{2}M_{3}M_{4})\ .\label{eq:LocalProb2}
\end{equation}

To calculate $M_{1}M_{2}M_{3}M_{4}$ we need to consider different
domains of $u$ separately to find different limiting properties of
the corner transfer matrices. Here we summarize the properties of
the corner matrices near the boundaries of domains ${\cal D}_{1}$
and ${\cal D}_{2}$ (see Appendix \ref{cmtprop} for details), which
are required for the calculation. In the domain ${\cal D}_{1}$, in
the $u\rightarrow0$ limit we find (upon dropping an irrelevant multiplicative
factor)

\begin{equation}
A(u=0)=C(u=0)=\openone,
\end{equation}
 while in the $u\rightarrow(2+r)\eta$ limit, we have

\begin{equation}
B(u=(2+r)\eta)=D(u=(2+r)\eta)=\tilde{V}{}_{1},
\end{equation}
 where we have defined 
\begin{equation}
(\tilde{V}{}_{1})_{\mathbf{l},\mathbf{l}'}=[h(2\eta l_{1})]^{1/2}\delta(\mathbf{l},\mathbf{l}').
\end{equation}
 Using the above limits for corner transfer matrices we can write
that 
\begin{equation}
A(u=0)B(u=(2+r)\eta)C(u=0)D(u=(2+r)\eta)=\tilde{V}{}_{1}^{2}.
\end{equation}
 Substituting Eq. (\ref{eq:CTMExp}) in the above equation we get
the following result for the product of the matrices $M_{1}M_{2}M_{3}M_{4}$,
which appear in the expression for the height probability in Eq. \eqref{eq:LocalProb2}
\begin{equation}
M_{1}M_{2}M_{3}M_{4}e^{2(2+r)\eta{\cal H}}=\tilde{V}{}_{1}^{2}\label{eq:ProdM1}
\end{equation}

In the domain ${\cal D}_{2}$, we can show that 
\begin{equation}
A(u=0)B(u=(2-r)\eta)C(u=0)D(u=(2-r)\eta)=\tilde{V}{}_{1}^{2}
\end{equation}
 and 
\begin{equation}
M_{1}M_{2}M_{3}M_{4}e^{2(2-r)\eta{\cal H}}=\tilde{V}{}_{1}^{2}.\label{eq:ProdM2}
\end{equation}
 Taking into account relations in Eqs. (\ref{eq:ProdM1}) and (\ref{eq:ProdM2}),
the local height probability can be written as 
\begin{equation}
P_{a}=Tr(S_{a}\tilde{V}{}_{1}^{2}e^{-2t\eta{\cal H}})/Tr(\tilde{V}{}_{1}^{2}e^{-2t\eta{\cal H}}),\label{eq:LocalHeight}
\end{equation}
 where 
\begin{equation}
t=\begin{cases}
2+r & u\in{\cal D}_{1}\\
2-r & u\in{\cal D}_{2}
\end{cases}.
\end{equation}

\subsection{Diagonal form of the corner transfer matrix $A$.}

What is left to do is to find the diagonal form of the corner transfer
matrix $A$ and the matrix ${\cal H}$, which is the most involved
part of the calculation. The corner transfer matrices in Eq. (\ref{eq:CTMExp})
should satisfy quasi-periodic conditions as do the weights, with the
period $2iK'$, which implies that elements of ${\cal H}$ are integer
multiples of $\pi/K'$, ${\cal H}_{\mathbf{l},\mathbf{l}'}=\pi N(\mathbf{l})\delta(\mathbf{l},\mathbf{l}')/K'$.
Similar to the solution of the RSOS models \cite{abf84}, we assume
that ${\cal H}$ does not change discontinuously with $p$, which
implies that the integer function $N(\mathbf{l})$ is independent
of $p$ and we can derive it in a limit where the composite weights
assume a simple form. We can show that (see Appendix \ref{weightprop})
in the $p\rightarrow1$ limit the weights of the composite model take
a particularly simple form and the corner transfer matrix $A$ can
be readily diagonalized. After a fairly lengthy calculation, involving
the `conjugate-modulus' (or modular) transformation, the dust settles,
and one finds the diagonal form of $A$ 
\begin{equation}
A_{\mathbf{l},\mathbf{l}'}=[e^{-u{\cal H}}]_{\mathbf{l},\mathbf{l}'}=g_{l_{1}}^{-1}w^{N(\mathbf{l})/2}\delta(\mathbf{l},\mathbf{l}')\ ,\label{eq:Adiagonal}
\end{equation}
 where 
\begin{align}
w & =e^{-2\pi u/K'} & g_{l_{1}} & =w^{(2l_{1}-r)^{2}/(16r)}
\end{align}
 and 
\begin{equation}
N(\mathbf{l})/2\equiv\phi(\bl)=\sum_{j=1}^{(m+1)/2}j\Bigl(\frac{|l_{2j+3}-l_{2j-1}|}{4}+\delta_{l_{2j-1},l_{2j+1}}\delta_{l_{2j+1},l_{2j+3}}\delta_{l_{2j},l_{2j+2}}\Bigr),\label{eq:phi-1}
\end{equation}
 where sum over $j$ is performed along a line in the 2D lattice (see
Fig. \ref{fig:CTM}). Each term in this sum corresponds to the weight
of the $j^{{\rm th}}$ plaquette (counted from the central site),
times $j$, which is the number of plaquettes on the $j^{{\rm th}}$
diagonal. In the limit $p=1$, these plaquettes all have the same
form, because $A$ is diagonal then, as shown in Appendix \ref{app:p-one-limit}.

\subsection{Local height probability.}

We can now collect the results, and obtain the local height probabilities
$P_{a}$. Substituting the diagonal form of $A$, Eq. \eqref{eq:Adiagonal}
in Eq. (\ref{eq:LocalHeight}), we find that

\begin{equation}
P_{a}=S^{-1}v_{a}X_{m}(a;b,c,d,e;x^{t}).
\end{equation}
 Here we adopted the following definitions:

\begin{equation}
X_{m}(a;b,c,d,e;q)=\sum_{l_{2},...,l_{m}}q^{\phi(\{l\})},\label{eq:Xfunction}
\end{equation}

\begin{equation}
v_{a}=x^{(2-t)(2a-r)^{2}/(16r)}E(x^{a},x^{r}),
\end{equation}

\begin{equation}
S=\sum_{a}v_{a}X_{m}(a;b,c,d,e;x^{t}),
\end{equation}

\begin{equation}
x=e^{-4\pi\eta/K'}\ .
\end{equation}
 The function $\phi(\bl)$ was given in Eq. \eqref{eq:phi-1} and
the function $E(z,x)$ appears in Jacobi's triple product identity,
\begin{equation}
E(z,x)=\prod_{n=1}^{\infty}(1-x^{n-1}z)(1-x^{n}z^{-1})(1-x^{n})\ .
\end{equation}
 Furthermore, $l_{1}=a$ and heights $b,c,d,e$ (see Fig. \ref{fig:CTM})
are the boundary heights, which are to be fixed to the values of a
ground state in the domain of $u$ under consideration. We see from
Eq. (\ref{eq:Xfunction}) that the partition function as well as the
local height probability is expressed as a sum over 1D height configurations,
in contrast to 2D configurations in the original formulation (see
Eq. (\ref{eq:PartitionFunction})). This property is the consequence
of integrability of the model and greatly simplifies calculations.

\section{Phases of the composite height model}

\label{sec:phases}

In this section, we will make a start with the exploration of the
phase diagram of the composite height model, which will be focussed
on the regions which are related to the anyonic quantum chain. We
will give a more detailed description of the various phases of the
model in a forthcoming publication.

The phase diagram of the composite height model bears resemblance
to the phase diagram of the original RSOS model. We will consider
the phase diagram as a function of the parameters $u$ and $p$, where
the parameter $u$ is related to the anisotropy of the model. Only
the sign of this parameter will be relevant. The parameter $p$ drives
a phase transition between different ordered and disordered phases,
as was the case in the RSOS model. We consider the regime $0\leq p\leq1$,
which is the one relevant for our purposes.

We start by analyzing the ordered phases by taking
either $u>0$, which we call regime III, or $u<0$, called regime
II. The naming of the regimes follows the nomenclature of ABF. We
use the results of the previous section, where we calculated the height
probabilities in terms of the functions 
\begin{equation}
X_{m}(a;b,c,d,e;q)=\sum_{l_{2},l_{3},\ldots,l_{m}}q^{\phi(\bl)}\ ,\label{eq:xm}
\end{equation}
 where the vector $\bl$ has $m+4$ components, $\bl=(a,l_{2},\ldots,l_{m},b,c,d,e)$,
which satisfy the constraint $l_{i}=l_{i-1}\pm1$, implying that $a+e=0\bmod2$.
We repeat the function $\phi(\bl)$ for convenience, 
\begin{equation}
\phi(\bl)=\sum_{j=1}^{(m+1)/2}j\Bigl(\frac{|l_{2j+3}-l_{2j-1}|}{4}+\delta_{l_{2j-1},l_{2j+1}}\delta_{l_{2j+1},l_{2j+3}}\delta_{l_{2j},l_{2j+2}}\Bigr)\ .\label{eq:phi}
\end{equation}
 For $0<p<1$, the height probabilities are proportional to $X_{m}(a;b,c,d,e;x^{t})$,
with $t=2+r$ for $u>0$, and $t=2-r$ for $u<0$.

To find the ground states, we analyze the function $\phi(\bl)$.
The ground states are those configurations which contribute maximally
to the partition function. In domain ${\cal D}_{1}$, which has $u>0$,
and $t=2+r>0$, the ground states are given by those configurations
which minimize the function $\phi(\bl)$. For domain ${\cal D}_{2}$,
with $u<0$ and $t=2-r<0$, the function $\phi(\bl)$ should be maximized
instead. As long as $0<p<1$, one finds that the arguments about the
ground states go through, because $x<1$. At the critical point ($p\rightarrow0$)
we have $x\rightarrow1$, hence the argument fails and all height
configurations contribute equally.

The model is critical when $p\rightarrow0$, and we will study the
full height probabilities $P_{a}$, which give the probability that
the central height takes the value $a$, depending on the boundary
heights $(b,c,d,e)$ (see Fig.~\ref{fig:CTM}). We will evaluate
these height probabilities, in the case that the boundary heights
are such that they belong to a ground state pattern.

\subsection{Ground states for $u>0$ (domain ${\cal D}_{1}$)}

\label{sec:gs-u-pos}

\begin{figure}
\includegraphics[scale=0.5]{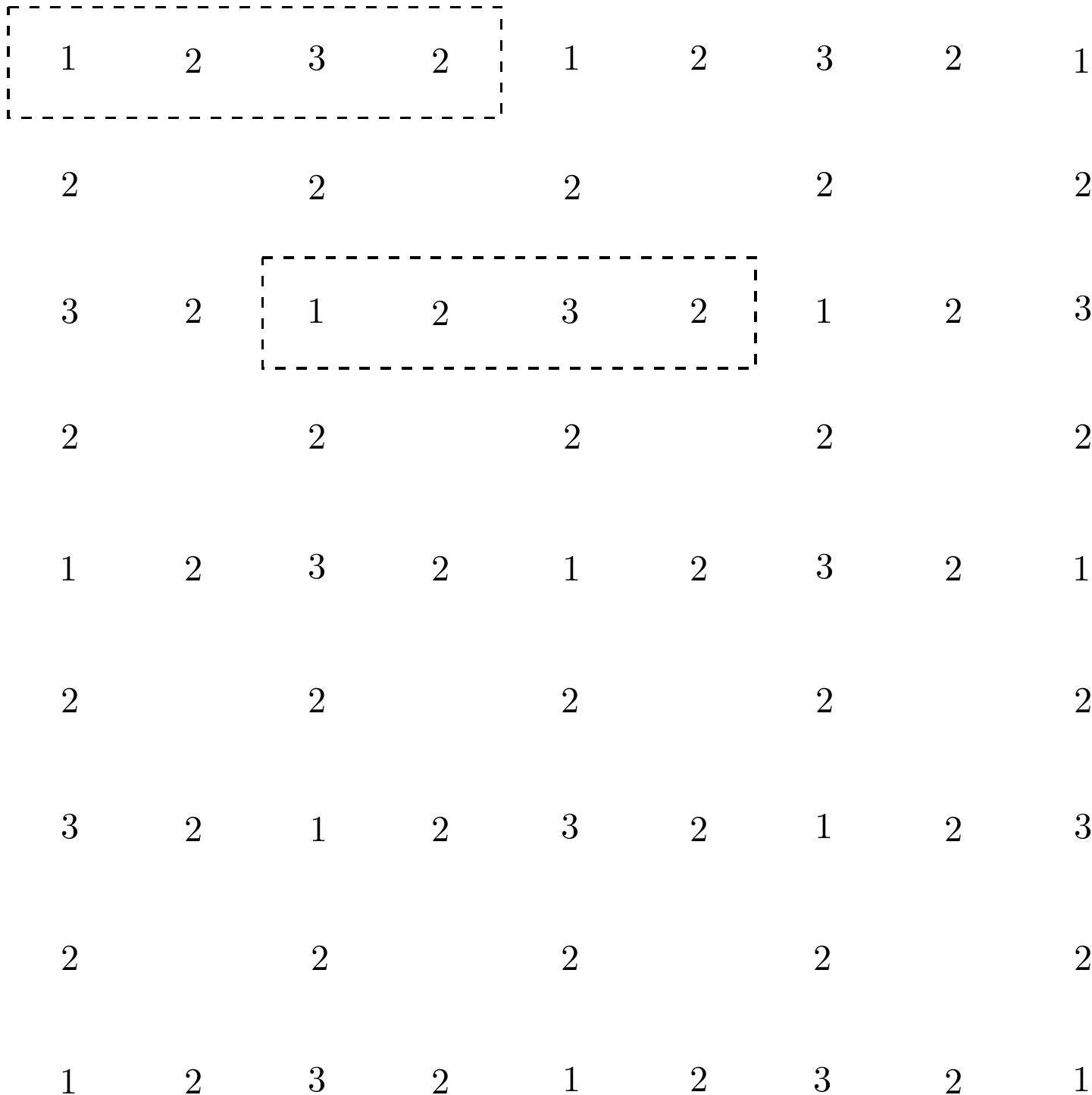} \caption{Ground state pattern for $u>0$, which is characterized by the pattern
in the dashed box.}

\label{fig:GroundState1} 
\end{figure}

Let us start by analyzing the case $u>0$, and minimize the function
$\phi(\bl)$. In this case, as many plaquettes as possible should
give zero contribution to $\phi(\bl)$. This can be achieved in the
following way. First of all, one should have that $l_{i}=l_{i+4}$,
such that the first term within the parenthesis in Eq. \eqref{eq:phi}
is zero. There are now two different ways in which one can avoid a
contribution from the second term. First, one can set $l_{i+2}=l_{i}\pm2$
and $l_{i+1}=l_{i+3}=l_{i}\pm1$, where the sign in both equations
should be the same. The other possibility is $l_{i+2}=l_{i}$, while
$l_{i+1}=l_{i}\pm1$ and $l_{i+3}=l_{i}\mp1$, where the signs in
the last two equations have to be opposite.

In particular, the vector $\bl$ describing the ground states for
$0<p<1$, $u>0$ takes the form (we show the case $r=7$) 
\begin{align*}
(1,2,3,2,1,2,3,2,1,\ldots) &  & (2,3,4,3,2,3,4,3,2,\ldots) &  & (3,4,5,4,3,4,5,4,3,\ldots) &  & (4,5,6,5,4,5,6,5,4,\ldots)\\
(3,2,1,2,3,2,1,2,3,\ldots) &  & (4,3,2,3,4,3,2,3,4,\ldots) &  & (5,4,3,4,5,4,3,4,5,\ldots) &  & (6,5,4,5,6,5,4,5,6,\ldots)
\end{align*}
 for the ground states of the first type. The patterns of the second
type are mere translations of the patterns of the first type, and
are given by the vectors $\bl$ of the form 
\begin{align*}
(2,1,2,3,2,1,2,3,2,\ldots) &  & (3,2,3,4,3,2,3,4,3,\ldots) &  & (4,3,4,5,4,3,4,5,4,\ldots) &  & (5,4,5,6,5,4,5,6,5,\ldots)\\
(2,3,2,1,2,3,2,1,2,\ldots) &  & (3,4,3,2,3,4,3,2,3,\ldots) &  & (4,5,4,3,4,5,4,3,4,\ldots) &  & (5,6,5,4,5,6,5,4,5,\ldots)
\end{align*}
 In the limit $p=1$, the corner transfer matrix $A$ is diagonal,
which gives rise to ground states patterns which are invariant under
translation along the North-East to South-West (NE-SW) diagonal. A
particular ground state pattern for $u>0$ is displayed in figure
\ref{fig:GroundState1}. All the ground states are of the form $(l-1,l,l+1,l,l-1,l,l+1,\ldots)$,
or translations of this pattern.

To count the number of different ground states, we note that the ground
state patterns are specified by three consecutive integers. Because
the heights can take the values $1,2,\ldots,r-1$, there are $r-3$
possible consecutive integers. By translation, each of these sets
of consecutive integers gives rise to four different ground states,
for a total of $4(r-3)$ ground states.

We will now count the number of different height probabilities, $P_{a}(b,c,d,e)$,
where the boundary condition $(b,c,d,e)$ corresponds to a ground
state pattern. First, we note that $a$, the height of the central
site, and $e$ have to have the same parity, $a+e=0\bmod2$. For $r$
odd, there are $(r-1)/2$ odd valued heights, and $(r-1)/2$ even
valued heights. In both cases, $a$ can take $(r-1)/2$ values, giving
$2(r-1)(r-3)$ different height probabilities. In the case that $r$
is even, there are $r/2$ odd valued heights, and $r/2-1$ even valued
heights. Out of the $4(r-3)$ ground state patterns, $2(r-3)$ have
$e$ even, and $2(r-3)$ have $e$ odd. Hence, also for $r$ even,
the number of height probabilities to consider is $2(r-1)(r-3)$.
These height probabilities are given in terms of the functions $X_{m}(a;b,c,d,e;q)$,
which satisfies the relation (which follows from the symmetry properties
of the plaquette weights) 
\begin{equation}
X_{m}(a;b,c,d,e;q)=X_{m}(r-a;r-b,r-c,r-d,r-e;q)\ .\label{eq:Xrelation}
\end{equation}
 This halves the number of independent height probabilities, which
is thus given by $(r-1)(r-3)$, or in terms of $k=r-2$, by $k^{2}-1$.

\subsection{Ground states for $u<0$ (domain ${\cal D}_{2}$)}

\label{sec:gs-u-neg}

\begin{figure}
\includegraphics[scale=0.5]{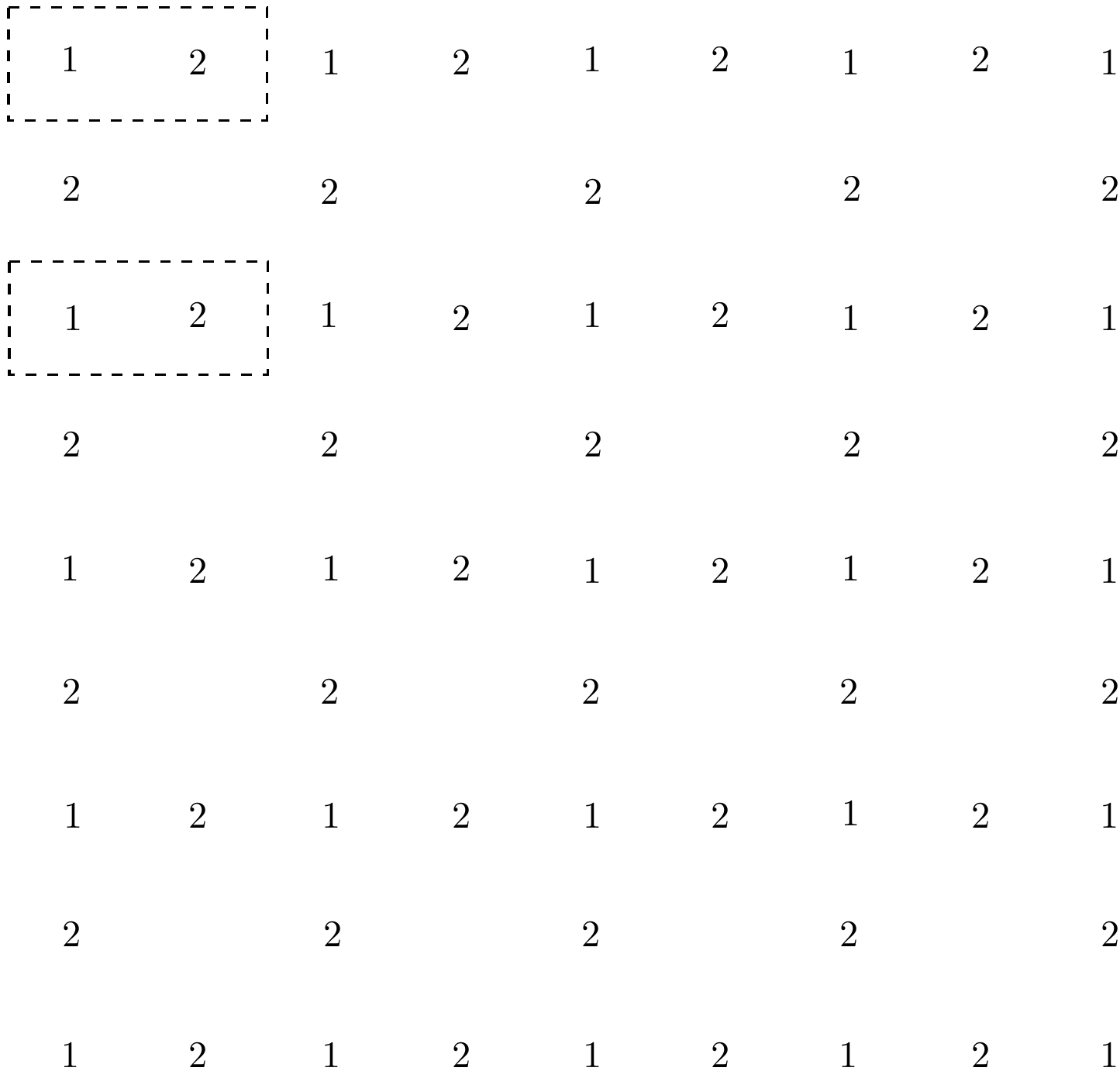} \caption{Ground state pattern for $u<0$, which is formed characterized by
the pattern in the dashed box.}

\label{fig:GroundState2} 
\end{figure}

We now consider the ground states for $u<0$, and look for configurations
which maximize the function $\phi(\bl)$. Because $l_{i}$ and $l_{i+4}$
maximally differ by four, both terms within the parentheses in Eq.
\eqref{eq:phi} can maximally contribute $1$. However, in order that
the first term contributes for every plaquette, the heights $l_{i}$
would have to steadily increase or decrease, which is impossible,
because the values the $l_{i}$ can take lie in the range $l_{i}=1,2,\ldots,r-1$.
The second term inside of the parenthesis in Eq. \eqref{eq:phi} can
be $1$ for all plaquettes. The only requirement is that $l_{2j-1}=l_{2j+1}=l_{2j+3}$
and $l_{2j}=l_{2j+2}=l_{2j-1}\pm1$. For $p=1$, the ground states
again are invariant under translations over the NE-SW diagonal, thus,
for $u<0$, the ground states are given by configurations for which
the heights stay as constant as possible, \emph{i.e.} they take the
form $(l,l+1,l,l+1,l,\ldots)$, as depicted in Fig. \ref{fig:GroundState2}
for a typical example.

The number of ground state patterns is given by the number of consecutive
pairs (both increasing and decreasing), \emph{i.e.} $2(r-2)$. We
again need the number of height probabilities we have to consider,
or the number of functions $X_{m}(a;b,c,d,e;q)$. We only have to
specify $(a;d,e)$, because $d$ and $e$ fix the values of $b$ and
$c$ in the ground states. There are $2(r-2)$ pairs $(d,e)$. It
turns out that the number of height probabilities is given by $(r-1)(r-2)$,
irrespective of whether $r$ is even or odd. Thus, there are $(r-1)(r-2)/2=k(k+1)/2$
independent height probabilities, because of the relation in Eq. \eqref{eq:Xrelation}.

\subsection{Criticality at $p=0$}

We found the ground states for $u>0$ and $u<0$ deep in the ordered
regime, namely for $0<p<1$, by minimizing or maximizing the function $\phi(\bl)$,
which appears in the height probabilities, as well as the partition
function. At the critical point, for $p=0$ as in the original RSOS
model, all configurations contribute, and one has to do a more careful
study of the model. In the remainder of this paper, we focus on the
full form of the height probabilities, which have a close connection
with conformal field theory characters, which allows us to unambiguously
identify the critical behavior of the model, for both regimes $u>0$
and $u<0$. Because we already established the connection between
the composite height model and the anyonic chains, we thereby also
explain the critical behavior of the anyonic chains at the integrable
point. A more detailed study of the critical behavior of the composite
height model will be left for a future publication.

\section{Evaluation of the height probabilities}

\label{sec:HeightProbabilities}

We now turn our attention to the evaluation of the height probabilities.
We have shown that the probability that the central height takes the
value $a$ depends on the boundary condition, which we specified by
fixing the boundary heights $(b,c,d,e)$. The height probabilities
are governed by the functions $X_{m}(a;b,c,d,e;q)$ given in Eqs.
\eqref{eq:xm} and \eqref{eq:phi}. We are ultimately interested in
the behavior of these functions in the limit $m\rightarrow\infty$.
In that limit, we can let the boundary heights correspond to the ground
state patterns, which extremize the function $\phi(\bl)$.

We therefore consider the functions $X_{m}(a;b,c,d,e;q)$, for all
possible values of $(a;b,c,d,e)$ such that the boundary heights $(b,c,d,e)$
are part of a ground state pattern. In the limit $m\rightarrow\infty$,
these functions will receive contributions from all possible configurations
specified by the vectors $\bl$. For finite $m$, these functions
are finite polynomials in $q$ (or $q^{1/2}$ times such polynomials).
The minimal power of $q$ which can occur is zero, while the maximal
power is $(m+1)(m+3)/8$. We note that these extremal values are not
obtained for all choices of $(a;b,c,d,e)$.

In the following subsections, we will provide explicit expressions
for the functions $X_{m}(a;b,c,d,e;q)$, in the case $r=5$. We did
not yet obtain explicit expressions for $r>5$, but by analyzing the
functions, we unambiguously identified them as the characters of certain
conformal field theories. In particular, we checked extensively that
in the limit of $m\rightarrow\infty$, the functions $X_{m}(q)$ tend
to affine Lie algebra branching functions, or in other words, characters
of the various primary fields in certain coset models. This is precisely
the connection we are after, because these coset models describe the
behavior of the model at the critical point, and hence the critical
behavior of the anyonic quantum chains. Before delving into the details,
we will first state the results here.

For $u>0$, we find that the functions $X_{m}(a;b,c,d,e;q)$ tend
to the characters of the coset $\frac{su(2)_{1}\times su(2)_{1}\times su(2)_{k-2}}{su(2)_{k}}$,
where $k=r-2$. In the case that $k=3$, this coset is equivalent
to a Gepner parafermion theory \cite{g87} based on $su(3)_{2}$,
explicitly $\frac{su(3)_{2}}{u(1)_{4}u(1)_{12}}$. The characters
of this theory were considered in Ref. \cite{arr01} in the context
of a particular non-abelian spin-singlet quantum Hall state \cite{as99}.
The finitizations of the characters considered in Ref.~\cite{arr01}
precisely correspond to the functions $X_{m}(q)$ we obtain from the
integrable model we introduced in this paper.

For $u<0$, one needs to consider the part of the functions $X_{m}(q)$
with the highest powers of $q$, as explained below. In particular,
the functions $q^{(m+1)(m+3)/8}X_{m}(q^{-1})$ are the functions one
needs to consider in the limit $m\rightarrow\infty$. We found that
they precisely correspond to the characters of the $Z_{k}$ parafermion
theory \cite{zf85} (which is for instance given by the coset $\frac{su(2)_{k}}{u(1)_{2k}}$).
The characters of this theory can be found in Ref. \cite{g95up}.

For $r=5$, we have the interesting situation that the finitization
of the characters of the $su(3)_{2}$ Gepner parafermions, \emph{i.e.}
the functions $X_{m}(q)$, also can be considered as finitized characters
of the $Z_{3}$ parafermions. Because both theories contain Fibonacci
particles, one could say that `Fibonacci meets Fibonacci', in the
same spirit as `Ising meets Fibonacci' \cite{gs09}, which establishes
a connection between the theories $su(2)_{2}$ and $su(3)_{2}$, which
have Ising and Fibonacci type fusion rules.

In the following two subsections, we provide explicit expressions
for the functions $X_{m}(a;b,c,d,e;q)$, for finite $m$ in the case,
$r=5$. These expressions are such that the limit $m\rightarrow\infty$
can be taken explicitly.

The identification of the functions $X_{m}(a;b,c,d,e;q)$ for all
values of $r$ and $(a;b,c,d,e)$ in terms of the CFT characters is
the subject of Appendix~\ref{app:cft-connection}, which deals with
both cases $u>0$ and $u<0$.

\subsection{Probability amplitudes for $u>0$}

In this subsection, we give the explicit form of the functions $X_{m}(a;b,c,d,e,q)$,
for those values of $(a;b,c,d,e)$ which correspond to the ground
states for $u>0$ (for $r=5$). We did not yet prove these results,
but we expect that a proof along the lines of the original paper \cite{abf84}
is feasible. Such a proof involves the recursion relations for the
functions $X_{m}(a;b,c,d,e,q)$, which we give in Appendix~\ref{app:cft-connection},
Eq.~\eqref{eq:Xrecursion}.

We start by giving the expressions for finite $m$, and take the limit
$m\rightarrow\infty$ afterwards. We introduce the function $(q)_{m}=\prod_{k=1}^{m}(1-q^{k})$,
for positive integers $m$, and $(q)_{0}=1$, which appears in the
definition of the $q$-binomials, or Gaussian polynomials 
\begin{equation}
\begin{bmatrix}m\\
n
\end{bmatrix}=\begin{cases}
\frac{(q)_{m}}{(q)_{n}(q)_{m-n}} & \text{for \ensuremath{0\leq n\leq m}}\\
0 & \text{otherwise}
\end{cases}
\end{equation}

The precise form of the following function we introduce was inspired
by the (finitized) character of the Gepner parafermions associated
with $su(3)_{2}$ (see Ref. \cite{arr01}), as well as the characters
for the original RSOS model (see, for instance, Ref. \cite{kkm93}).
In particular, we introduce 
\begin{equation}
y(m;l_{2},l_{3},l_{4};q)=\sideset{}{'}\sum_{a,b\geq0}q^{\frac{a^{2}+b^{2}-ab-a\delta_{l_{4},3}-b\delta_{l_{4},2}}{2}}\begin{bmatrix}\frac{m+b+\delta_{l_{3},1}+\delta_{l_{4},3}}{2}\\
a
\end{bmatrix}\begin{bmatrix}\frac{m+a+\delta_{l_{2},1}+\delta_{l_{3},2}+\delta_{l_{4},2}}{2}\\
b
\end{bmatrix}\ .
\end{equation}
 where we assume that $m$ is an integer, $l_{2},l_{3}=1,2$ and $l_{4}=1,2,3,4$.
The prime on the sum indicates the constraints that the argument of
the $q$-binomials have to be (non-negative) integers. We labelled
the function with $l_{2},l_{3},l_{4}$, because of the connection
with conformal field theory characters, which we will describe below
(the label $l_{1}$ can always be chosen as $l_{1}=1$).

We then have the following results 
\begin{equation}
\begin{split}X_{m}(1;2,1,2,3;q) & =y(\frac{m-1}{2};1,1,1;q)\\
X_{m}(3;2,1,2,3;q) & =y(\frac{m-1}{2};1,1,3;q)\\
X_{m}(2;3,2,3,4;q) & =y(\frac{m-1}{2};1,2,2;q)\\
X_{m}(4;3,2,3,4;q) & =y(\frac{m-1}{2};1,2,4;q)\\
X_{m}(2;1,2,3,2;q) & =y(\frac{m-1}{2};2,1,2;q)\\
X_{m}(4;1,2,3,2;q) & =y(\frac{m-1}{2};2,1,4;q)\\
X_{m}(1;2,3,4,3;q) & =y(\frac{m-1}{2};2,2,1;q)+q^{\frac{m+1}{4}}y(\frac{m-3}{2};1,1,1;q)\\
X_{m}(3;2,3,4,3;q) & =y(\frac{m-1}{2};2,2,3;q)+q^{\frac{m+1}{4}}y(\frac{m-3}{2};1,1,3;q)
\end{split}
\label{eq:xmnegu}
\end{equation}

We can now rather easily take the limit $m\rightarrow\infty$, by
using the result $\lim_{n\rightarrow\infty}\begin{bmatrix}n\\
n'
\end{bmatrix}=\frac{1}{(q)_{n'}}$. We will assume that $m$ is of the form $4p+3$, with $p$ integer
(the case $m=4p+1$ is only slightly different). By taking this limit,
we find that 
\begin{align}
\lim_{p\rightarrow\infty}X_{4p+3}(1;2,1,2,3;q) & =c_{\id}^{{\rm su}(3)_{2}}(q)\nonumber \\
\lim_{p\rightarrow\infty}X_{4p+3}(4;3,2,3,4;q) & =\lim_{p\rightarrow\infty}X_{4p+3}(4;1,2,3,2;q)=\lim_{p\rightarrow\infty}X_{4p+3}(1;2,3,4,3;q)=c_{\psi}^{{\rm su}(3)_{2}}(q)\\
\lim_{p\rightarrow\infty}X_{4p+3}(2;3,2,3,4;q) & =\lim_{p\rightarrow\infty}X_{4p+3}(2;1,2,3,2;q)=\lim_{p\rightarrow\infty}X_{4p+3}(3;2,3,4,3;q)=q^{-\frac{1}{10}}c_{\sigma}^{{\rm su}(3)_{2}}(q)\nonumber \\
\lim_{p\rightarrow\infty}X_{4p+3}(3;2,1,2,3;q) & =q^{-\frac{1}{10}}c_{\rho}^{{\rm su}(3)_{2}}(q)\nonumber 
\end{align}
 Here, the functions $c_{x}^{{\rm su}(3)_{2}}(q)$ denote the characters
of the ${\rm su}(3)_{2}$ parafermion theory. This theory has eight
fields, including the identity field $\id$, with the character $c_{\id}^{{\rm su}(3)_{2}}(q)$,
and three parafermions, $\psi_{1}$, $\psi_{2}$ and $\psi_{12}$,
which have identical characters $c_{\psi}^{{\rm su}(3)_{2}}(q)$.
The remaining four fields are three `spin' fields $\sigma_{1}$, $\sigma_{2}$
and $\sigma_{12}$, which have identical characters $c_{\sigma}^{{\rm su}(3)_{2}}(q)$.
Finally there is the field $\rho$, whose character we denote by $c_{\rho}^{{\rm su}(3)_{2}}(q)$.

In the above, the characters of the fields are `normalized' such that
the first term in the expansion is $q^{h}$, where $h$ is the conformal
dimension of the field under consideration. These conformal dimensions
are given by $h_{\id}=0$, $h_{\psi}=\frac{1}{2}$, $h_{\sigma}=\frac{1}{10}$,
and $h_{\rho}=\frac{3}{5}$. Combining all the factors, the expressions
for the limit of $X_{m}(q)$ are series expansions with integer or
half integer powers of $q$.

For our present proposes, it is best to view this CFT as the following
coset model: $\frac{su(2)_{1}\times su(2)_{1}\times su(2)_{1}}{su(2)_{3}}$.
The fields in this coset model are labelled by the labels of the constituent
factors, $\Phi_{l_{4}}^{l_{1},l_{2},l_{3}}$, where $l_{1},l_{2},l_{3}$
correspond to the factors $su(2)_{1}$, and $l_{4}$ corresponds to
the factor $su(2)_{3}$. These labels have to satisfy the constraint
$l_{1}+l_{2}+l_{3}+l_{4}=0\bmod2$, which we use to set $l_{1}=0$,
and consider the fields with $l_{2}+l_{3}+l_{4}=0\bmod2$.

For completeness, we explicitly give the labels of the parafermion
fields 
\begin{align*}
\Phi_{0}^{0,0,0} & =\id & \Phi_{0}^{0,1,1} & =\psi_{1} & \Phi_{3}^{0,1,0} & =\psi_{2} & \Phi_{3}^{0,0,1} & =\psi_{12}\\
\Phi_{2}^{0,0,0} & =\rho & \Phi_{2}^{0,1,1} & =\sigma_{2} & \Phi_{1}^{0,1,0} & =\sigma_{1} & \Phi_{1}^{0,0,1} & =\sigma_{12}\ .
\end{align*}

For $r>5$, we did not yet obtain closed expressions for the functions
$X_{m}(a;b,c,d,e;q)$. However, one can obtain expansions to high
order, by making use of the recursion relations satisfied by the $X_{m}(q)$,
Eq.~\eqref{eq:Xrecursion}. These high order expansions can be compared
to the branching functions (or characters) of various coset model.
In doing so, we have established that the $X_{m}(a;b,c,d,e;q;r)$
are (in the limit $m\rightarrow\infty$) the characters of the cosets
$\frac{su(2)_{1}\times su(2)_{1}\times su(2)_{k-2}}{su(2)_{k}}$ (we
remind that $k=r-2$). In Appendix~\ref{app:cft-connection}, we
will give the relation between the values $(a;b,c,d,e)$ and the labels
of the coset fields.

\subsection{Probability amplitudes for $u<0$}

The functions $X_{m}(q)$, with the boundary heights $(b,c,d,e)$
given by the ground state patterns for $u<0$, can also be expressed
in terms of the function $y(m;l_{2},l_{3},l_{4})$ which was introduced
in the previous subsection. We first state these results, and subsequently
take the limit $m\rightarrow\infty$, in order to identify the critical
theory describing the critical behavior of the anyonic quantum spin
chain. In particular, for $r=5$ 
\begin{equation}
\begin{split}X_{m}(1;2,1,2,1;q) & =q^{\frac{m+1}{4}}y(\frac{m-1}{2},1,1,1)\\
X_{m}(1;2,3,2,3;q) & =q^{\frac{m+1}{4}}y(\frac{m-1}{2},1,2,1)\\
X_{m}(1;4,3,4,3;q) & =q^{\frac{m+1}{2}}y(\frac{m-1}{2},2,2,1)\\
X_{m}(2;1,2,1,2;q) & =q^{\frac{m+1}{2}}y(\frac{m-1}{2},2,2,2)\\
X_{m}(2;3,2,3,2;q) & =q^{\frac{m+1}{4}}y(\frac{m-1}{2},1,2,3)\\
X_{m}(2;3,4,3,4;q) & =q^{\frac{m+1}{4}}y(\frac{m-1}{2},1,1,3)
\end{split}
\end{equation}
 The ground states correspond to the highest possible powers of $q$,
so to make the identification with the conformal field theory, we
we will have to make the substitution $q\rightarrow q^{-1}$, and
multiply by $q^{\frac{(m+1)(m+3)}{8}}$, which is the maximal power
of $q$ which occurs in the functions $X_{m}(q)$. This will make
sure that the function corresponding to the vacuum character is a
polynomial in $q$, starting with $1$. We find that the functions
thus obtained tend to the characters of the $su(2)_{3}$ parafermions
in the limit $m\rightarrow\infty$. The $su(2)_{3}$ parafermion theory
contains six fields, the identity, with character $c_{\id}^{{\rm su(2)_{3}}}$,
two parafermion fields $\psi_{1}$ and $\psi_{2}$, with the character
$c_{\psi}^{{\rm su(2)_{3}}}$, two spin fields $\sigma_{1}$ and $\sigma_{2}$,
with characters $c_{\sigma}^{{\rm su(2)_{3}}}$, and finally the field
$\epsilon$, with the character $c_{\epsilon}^{{\rm su(2)_{3}}}$.
The scaling dimensions are $h_{\id}=0$, $h_{\psi}=\frac{2}{3}$,
$h_{\sigma}=\frac{1}{15}$ and $h_{\epsilon}=\frac{2}{5}$. We find
\begin{align}
\lim_{p\rightarrow\infty}q^{\frac{(p+1)(p+2)}{2}}X_{2p+1}(1;2,1,2,1;q^{-1}) & =c_{\id}^{{\rm su}(2)_{3}}(q)\\
\lim_{p\rightarrow\infty}q^{\frac{(p+1)(p+2)}{2}}X_{2p+1}(1;2,3,2,3;q^{-1}) & =\lim_{p\rightarrow\infty}q^{\frac{(p+1)(p+2)}{2}}X_{2p+1}(1;4,3,4,3;q^{-1})=q^{-\frac{1}{6}}c_{\psi}^{{\rm su}(2)_{3}}(q)\\
\lim_{p\rightarrow\infty}q^{\frac{(p+1)(p+2)}{2}}X_{2p+1}(2;1,2,1,2;q^{-1}) & =\lim_{p\rightarrow\infty}q^{\frac{(p+1)(p+2)}{2}}X_{2p+1}(2;3,2,3,2;q^{-1})=q^{-\frac{1}{15}}c_{\sigma}^{{\rm su}(2)_{3}}(q)\\
\lim_{p\rightarrow\infty}q^{\frac{(p+1)(p+2)}{2}}X_{2p+1}(2;3,4,3,4;q^{-1}) & =q^{\frac{1}{10}}c_{\epsilon}^{{\rm su}(2)_{3}}(q)
\end{align}
 The identification of the functions $X_{m}(q^{-1})$ for $u<0$ in
the limit $m\rightarrow\infty$ is given in Appendix \ref{app:cft-connection},
in the case $k>3$ ($r>5$). They correspond to the characters of
the $Z_{k}$ parafermions \cite{g95up}.

\section{Conclusions and Outlook}

\label{sec:conclusions}

We have introduced a two-dimensional classical statistical mechanics
model the critical properties of which correspond to the integrable
points of a chain of $su(2)_{k}$ anyons with competing two- and tree-body
interactions ($J_{2}-J_{3}$ model, given in Eq.~(\ref{eq:ham-int})).
The 2D classical model is a composite height model which is a generalization
of the restricted solid-on-solid model solved by Andrews, Baxter and
Forrester \cite{abf84} by means of the corner transfer matrix method.
We have also used the CTM method and have found that, similar to the
RSOS model, there are four different regimes with two critical points
at $p=0$. We have studied two new integrable critical points of the
anyonic chain at $\tan\theta_{{\rm int}}=(d_{k}^{2}-1)/d_{k}^{2}$,
which correspond to the $p\rightarrow+0$, $u\rightarrow\pm0$ limits
of the composite height model.

For ferromagnetic interactions ($u>0$, $\theta=\arctan((d_{k}^{2}-1)/d_{k}^{2})+\pi$,
$J_{2},J_{3}<0$) the critical point is described by the $\frac{su(2)_{1}\times su(2)_{1}\times su(2)_{k-2}}{su(2)_{k}}$
coset conformal field theory. This critical behavior actually describes
an extended critical region around the integrable point. For antiferromagnetic
interactions ($u<0$, $\theta=\arctan((d_{k}^{2}-1)/d_{k}^{2})$,
$J_{2},J_{3}>0$) the behavior is that of the $Z_{k}$ parafermions.
This critical point constitutes the boundary between a gapped phase,
and an extended critical region, which is described, in general, by
the $\mathcal{M}(k+1,k+2)$ minimal model \cite{ftl07}.

These CFT identifications stem from the observations that functions
$X_{m}(q)$, which define the local height probabilities are given
by characters of the corresponding conformal field theory. The integrable
properties of quantum 1D and classical 2D models are fully defined
by the Temperley-Lieb algebra relations in Eq. (\ref{eq:TLAlgebra}),
which suggest that the Hamiltonian in Eq. (\ref{eq:ham-int}) can
be exactly solved in different representations of TL algebra. A particular
physical interpretation of the model depends on the particular representation
chosen to solve it. In our case, we have used the representation of
$su(2)_{k}$ anyons, which straightforwardly maps onto the composite
RSOS model. In this representation, an interpretation as a chain of
interacting anyons is straightforward.

A different interpretation as a $Q$-state Potts model (with $\sqrt{Q}=2\cos(\frac{\pi}{k+2})=d_{k}$,
where in this case, $k$ is considered to be a continuous parameter)
or a six-vertex model was put forward by Ikhlef \textit{et. al.} \cite{ijs09,ijs10},
which naturally admits the so called `loop representation'. The $J_{2}-J_{3}$
model at the integrable point, $\theta=\arctan((d_{k}^{2}-1)/d_{k}^{2})+\pi$,
has been exactly solved in spin-1/2 representation of the $U_{q}(SU(2))$
quantum algebra \cite{ijs09,ijs10}. In this representation, in contrast
to the `anyon representation', the Hilbert space has a tensor product
structure and the $J_{2}-J_{3}$ model admits the solution by the
algebraic Bethe Ansatz method. We should note that in spin or loop
representations the $1D$ quantum Hamiltonian is non-Hermitian and
lacks obvious physical interpretation. Despite this non-Hermiticity,
it has been conjectured that the part of the spectrum which scales
as $1/L$ is real. In contrast, in the `anyon representation', which
is adopted in this paper, the Hamiltonian is Hermitian (real symmetric)
and has a physical interpretation as a local interaction between anyons.
Despite the differences, the critical properties at $\theta=\arctan((d_{k}^{2}-1)/d_{k}^{2})+\pi$
found by us are similar to those found by Ikhlef \textit{et. al.}
\cite{ijs09,ijs10}. Namely, they have the same central charge, at
least for $u$ positive. The precise connection between these models
is interesting, and requires more detailed research. We note that
a similar loop model, related to the anyon model with $u<0$ ($\theta=\arctan((d_{k}^{2}-1)/d_{k}^{2})$
), has been studied in Ref.~\cite{js06}. It was conjectured and
several arguments were given that the critical behavior of this loop
model (in an appropriate limit) is described by $Z_{k}$ parafermions.

The anyonic chains have been generalized in different ways. First,
to different type of unitary anyons \cite{gat09}. It would be very
interesting to investigate to what extend the methods of `fused models',
as studied in Refs. \cite{djk87} and \cite{ks93} can be applied
to the anyonic chains of Ref. \cite{gat09}. Chains of non-unitary
anyons were also considered \cite{agl11}. This leads in general to
non-Hermitian Hamiltonians, which nevertheless have a real spectrum.
It should be possible to generalize the models considered in Ref.
\cite{fb85} to composite versions, and make a connection with a generalization
of the chains considered in Ref. \cite{agl11}.

In this paper, we have described only half of the phase diagram of
the composite RSOS model, positive $p$ regime ($0<p<1$). The corner
transfer matrix method allows to study the negative $p$ regime ($-1<p<0$)
also, which will be addressed in a subsequent publication. We have
shown that the positive $p$ regime has the interpretation in terms
of 1D interacting anyons. The 1D quantum mechanical interpretation
of the negative $p$ regime is yet unknown and also requires further
investigation.

Last, but not least, we would like to point out that the study of
the RSOS model by Andrews, Baxter and Forrester gave rise to an interesting
set of Rogers-Ramanujan-type identities. Further study of the composite
height model in light of these identities will be most interesting.

\noindent {\em Acknowledgements.} We acknowledge stimulating discussions
with C.J. Bolech, H.P. Eckle, Y. Ikhlef, J. Jacobson, A.W.W. Ludwig
and S. Trebst. We thank K. Schoutens for helpful comments on the literature,
and M. Hermanns, Y. Ikhlef and S. Trebst for useful comments on the
manuscript. P.K. acknowledges support from the NSF grant DMR-1006684.

\appendix

\section{The plaquettes of the composite-RSOS model\label{sec:plaquettes}}

\label{app:plaquettes}

The plaquettes of the composite model are obtained by stacking four
plaquettes of the original RSOS model (see Fig.~\ref{fig:rsos-weights}),
and giving them the appropriate weights. In the process of stacking
the plaquettes, one has to sum over the internal height, which in
some cases can take two possible values (see Fig.~\ref{fig:Gen-plaquette}).

In the following, we will display the possible composite plaquette
types, and give the associated weights in terms of the weights of
the RSOS model. In this case, the composite plaquette weights have
the same symmetry as the original model (with $\phi=0$), namely the
weights are unchanged under exchange of the North-West (NW) and South-East
(SE) corners, as well as under exchange of the North-East (NE) and
South-West (SW) corners. This last property is lost, if $\phi\neq0,K$.

We start by giving the composite plaquettes for which the internal
height is fixed by the boundary heights. In case the plaquette is
not symmetric, we will indicate the amount of plaquettes which can
be obtained from the given by taking the appropriate mirror image.
These mirrored plaquettes have the same weight as the displayed plaquette.

The weights of the composite plaquettes are given in terms of the
weights of the original RSOS model, given in Eq. \eqref{eq:rsos-weights}.
Let $\zeta_{l}(u)$ be any weight of the RSOS model, \emph{i.e.} $\zeta=\alpha,\beta,\gamma,\delta$.
Then, $\zeta_{l}=\zeta_{l}(u)$, $\zeta_{l}^{+}=\zeta_{l}(u+K)$ and
$\zeta_{l}^{-}=\zeta_{l}(u-K)$. With this notation, we can give the
weights of the composite plaquettes for $\phi=K$:

\begin{align}
\,\,\,\compp{l+2}{l+1}{l+1}{l}{}{l}{l-1}{l-1}{l-2} & =\beta_{l+1}^{+}\beta_{l-1}^{-}\beta_{l}^{2}\quad(\nesw) & \compp{l}{l+1}{l-1}{l+2}{}{l-2}{l+1}{l-1}{l} & =\alpha_{l}^{+}\alpha_{l}^{-}\alpha_{l+1}\alpha_{l-1}\quad(\nwse)\\
\nonumber \\
\compp{l+1}{l}{l}{l-1}{}{l-1}{l-2}{l-2}{l-1} & =\beta_{l}^{+}\gamma_{l-2}^{-}\beta_{l-1}^{2}\quad(\nesw) & \compp{l-1}{l}{l}{l+1}{}{l+1}{l+2}{l+2}{l+1} & =\beta_{l}^{+}\delta_{l+2}^{-}\beta_{l+1}^{2}\quad(\nesw)\\
\nonumber \\
\compp{l-1}{l}{l-2}{l+1}{}{l-1}{l}{l-2}{l-1} & =\alpha_{l-1}^{+}\alpha_{l-1}^{-}\alpha_{l}\delta_{l-1}\quad(\nwse) & \compp{l+1}{l}{l+2}{l-1}{}{l+1}{l}{l+2}{l+1} & =\alpha_{l+1}^{+}\alpha_{l+1}^{-}\alpha_{l}\gamma_{l+1}\quad(\nwse)\\
\nonumber \\
\compp{l+1}{l}{l}{l-1}{}{l-1}{l-2}{l}{l-1} & =\beta_{l}^{+}\alpha_{l-1}^{-}\beta_{l-1}\gamma_{l-1}\quad(\symd) & \compp{l-1}{l}{l}{l+1}{}{l+1}{l+2}{l}{l+1} & =\beta_{l}^{+}\alpha_{l+1}^{-}\beta_{l+1}\delta_{l+1}\quad(\symd)\\
\nonumber \\
\compp{l-1}{l}{l}{l+1}{}{l-1}{l}{l-2}{l-1} & =\delta_{l}^{+}\alpha_{l-1}^{-}\alpha_{l}\beta_{l-1}\quad(\symd) & \compp{l+1}{l}{l}{l-1}{}{l+1}{l}{l+2}{l+1} & =\gamma_{l}^{+}\alpha_{l+1}^{-}\alpha_{l}\beta_{l+1}\quad(\symd)\\
\nonumber \\
\compp{l+1}{l}{l}{l-1}{}{l+1}{l-2}{l}{l-1} & =\beta_{l}^{+}\alpha_{l-1}^{-}\beta_{l-1}\alpha_{l}\quad(\symd) & \compp{l-1}{l}{l}{l+1}{}{l-1}{l+2}{l}{l+1} & =\beta_{l}^{+}\alpha_{l+1}^{-}\beta_{l+1}\alpha_{l}\quad(\symd)\\
\nonumber \\
\compp{l}{l+1}{l+1}{l}{}{l}{l+1}{l-1}{l} & =\delta_{l+1}^{+}\alpha_{l}^{-}\gamma_{l}\beta_{l}\quad(\symd) & \compp{l}{l-1}{l-1}{l}{}{l}{l-1}{l+1}{l} & =\gamma_{l-1}^{+}\alpha_{l}^{-}\delta_{l}\beta_{l}\quad(\symd)\\
\nonumber \\
\compp{l}{l+1}{l+1}{l}{}{l}{l-1}{l-1}{l} & =\delta_{l+1}^{+}\gamma_{l-1}^{-}\beta_{l}^{2}\quad(\nesw) & \compp{l}{l+1}{l-1}{l}{}{l}{l-1}{l+1}{l} & =\alpha_{l}^{+}\alpha_{l}^{-}\beta_{l}^{2}\quad(\nwse)\\
\nonumber \\
\compp{l}{l+1}{l-1}{l}{}{l}{l+1}{l-1}{l} & =\alpha_{l}^{+}\alpha_{l}^{-}\gamma_{l}\delta_{l}\quad(\nwse)
\end{align}

We now focus on the composite plaquettes for which the internal height
is not fixed by those of the boundary in general, and hence takes
two different values. The composite weight is a sum of two terms,
the first term correspond to the internal height being $l+1$, the
second one to the internal height $l-1$:

\begin{align}
\compp{l+1}{l}{l}{l+1}{}{l+1}{l}{l}{l+1} & =\gamma_{l}^{+}\gamma_{l}^{-}\delta_{l+1}^{2}+\beta_{l}^{+}\beta_{l}^{-}\alpha_{l}^{2} & \compp{l-1}{l}{l}{l-1}{}{l-1}{l}{l}{l-1} & =\beta_{l}^{+}\beta_{l}^{-}\alpha_{l}^{2}+\delta_{l}^{+}\delta_{l}^{-}\gamma_{l-1}^{2}\\
\nonumber \\
\compp{l-1}{l}{l}{l+1}{}{l+1}{l}{l}{l+1} & =\beta_{l}^{+}\gamma_{l}^{-}\delta_{l+1}^{2}+\delta_{l}^{+}\beta_{l}^{-}\alpha_{l}^{2}\quad(\nesw) & \compp{l+1}{l}{l}{l-1}{}{l-1}{l}{l}{l-1} & =\gamma_{l}^{+}\beta_{l}^{-}\alpha_{l}^{2}+\beta_{l}^{+}\delta_{l}^{-}\gamma_{l-1}^{2}\quad(\nesw)\label{eq:off-diagonal1}\\
\nonumber \\
\compp{l+1}{l}{l}{l-1}{}{l+1}{l}{l}{l+1} & =\gamma_{l}^{+}\gamma_{l}^{-}\delta_{l+1}\alpha_{l}+\beta_{l}^{+}\beta_{l}^{-}\alpha_{l}\gamma_{l-1}\quad(\nwse) & \compp{l-1}{l}{l}{l+1}{}{l-1}{l}{l}{l-1} & =\beta_{l}^{+}\beta_{l}^{-}\delta_{l+1}\alpha_{l}+\delta_{l}^{+}\delta_{l}^{-}\alpha_{l}\gamma_{l-1}\quad(\nwse)\\
\nonumber \\
\compp{l-1}{l}{l}{l+1}{}{l+1}{l}{l}{l-1} & =\beta_{l}^{+}\beta_{l}^{-}\delta_{l+1}^{2}+\delta_{l}^{+}\delta_{l}^{-}\alpha_{l}^{2} & \compp{l+1}{l}{l}{l-1}{}{l-1}{l}{l}{l+1} & =\gamma_{l}^{+}\gamma_{l}^{-}\alpha_{l}^{2}+\beta_{l}^{+}\beta_{l}^{-}\gamma_{l-1}^{2}\\
\nonumber \\
\compp{l-1}{l}{l}{l-1}{}{l+1}{l}{l}{l+1} & =\beta_{l}^{+}\gamma_{l}^{-}\alpha_{l}\delta_{l+1}+\delta_{l}^{+}\beta_{l}^{-}\gamma_{l-1}\alpha_{l}\quad(\symd)\label{eq:off-diagonal2}
\end{align}

We find that the total number of possible composite plaquette types
is $66$. The total number of plaquettes depends on the value of $r=k+2$,
as is the case in the RSOS model.

\section{Properties of corner transfer matrices in different domains\label{cmtprop}}

\label{app:ctm-properties}

To find out the properties of corner transfer matrices near the boundaries
of different domains we need to calculate the weights of the composite
model in those limits.

\subsection{$u\rightarrow0$ limit}

Using the properties of the elliptic theta functions it is straightforward
to show that the weights of the RSOS model in the $u\rightarrow0$
limit are

\begin{eqnarray}
\alpha_{l}(u=0) & = & 1,\\
\beta_{l}(u=0) & = & 0,\nonumber \\
\gamma_{l}(u=0) & = & 1,\nonumber \\
\delta_{l}(u=0) & = & 1.\nonumber 
\end{eqnarray}

The RSOS wights with shifted fugacities, $\phi=K=\eta r$, have the
form

\begin{eqnarray}
\alpha_{l}^{+}(u=0) & = & -\alpha_{l}^{-}(u=0)=\frac{h(2\eta-\eta r)}{h(2\eta)},\label{eq:u0-relations}\\
\beta_{l}^{+}(u=0) & = & -\beta_{l}^{-}(u=0)=\frac{h(\eta r)}{h(2\eta)}\frac{[h(2\eta(l-1))h(2\eta(l+1))]^{1/2}}{h(2\eta l)},\nonumber \\
\gamma_{l}^{+}(u=0) & = & -\gamma_{l}^{-}(u=0)=\delta_{l}^{-}(u=0)=\frac{h(2\eta l+\eta r)}{h(2\eta l)},\nonumber \\
\delta_{l}^{+}(u=0) & = & -\delta_{l}^{-}(u=0)=\gamma_{l}^{-}(u=0)=\frac{h(2\eta l-\eta r)}{h(2\eta l)}.\nonumber 
\end{eqnarray}

We are able to show that in this limit all off-diagonal (NE-SW asymmetric)
weights of the composite model are zero and only diagonal (NE-SW symmetric)
weights survive. This implies that the corner transfer matrices $A$
and $C$ are diagonal in this limit. Of the 66 different type of plaquettes,
50 are off-diagonal. Of these 50 type of plaquettes, 42 are trivially
zero in the limit $u\rightarrow0$, because they contain a factor
$\beta_{l}(u=0)=0$. The remaining plaquettes are those in Eqs.~\eqref{eq:off-diagonal1}
and \eqref{eq:off-diagonal2}, and they are zero because of the relations
between the weights given in Eq.~\eqref{eq:u0-relations}. As an
example, we have for the plaquettes on left hand side of Eq.~\eqref{eq:off-diagonal1}:
$\beta_{l}^{+}\gamma_{l}^{-}\delta_{l+1}^{2}+\delta_{l}^{+}\beta_{l}^{-}\alpha_{l}^{2}=\beta_{l}^{+}\gamma_{l}^{-}+\delta_{l}^{+}\beta_{l}^{-}=\beta_{l}^{+}\delta_{l}^{+}-\delta_{l}^{+}\beta_{l}^{+}=0$.

The weights of all the diagonal plaquettes turn out to be the same.
For the weights which only have one contributing term, we immediately
find $W_{1}=-\bigl(\frac{h(2\eta-\eta r)}{h(2\eta)}\bigr)^{2}$. The
weights of the diagonal plaquettes which consist of two terms read
$W_{2}=-\bigl(\frac{h(2\eta l-\eta r)}{h(2\eta l)}\bigr)^{2}-\bigl(\frac{h(\eta r)}{h(2\eta)h(2\eta l)}\bigr)^{2}h(2\eta l-2\eta)h(2\eta l+2\eta)$.
By making use of the following identity for elliptic functions (see,
for instance, Chapter 15 of Ref.~\cite{book:b82}) 
\begin{equation}
h^{2}(2\eta-\eta r)h^{2}(2\eta l)=h^{2}(2\eta)h^{2}(2\eta l-\eta r)+h^{2}(\eta r)h(2\eta(l-1))h(2\eta(l+1))\ ,\label{eq:elliptic-relation}
\end{equation}
 we find that $W_{2}=W_{1}$. Thus, the general weight of the composite
model reads, in the limit $u\rightarrow0$ 
\begin{equation}
\tilde{W}(l_{1},l_{2},l_{3,}l_{4},l_{5},l_{6},l_{7},l_{8})(u=0)=-\left(\frac{h(2\eta-\eta r)}{h(2\eta)}\right)^{2}\delta_{l_{2},l_{8}}\delta_{l_{3},l_{7}}\delta_{l_{4},l_{6}}\ .
\end{equation}
 Using the definition of $U_{j}$, Eq. (\ref{eq:defU}), we can show
that 
\begin{equation}
(U_{j})_{\mathbf{l},\mathbf{l}'}(u=0)=-\left(\frac{h(2\eta-\eta r)}{h(2\eta)}\right)^{2}\delta(\mathbf{l},\mathbf{l}')\ ,
\end{equation}
 such that $U_{j}$ is a diagonal matrix. From the above and the Eq.
(\ref{eq:cmtA}) it follows that 
\begin{equation}
A(u=0)=C(u=0)=\openone,
\end{equation}
 where we have dropped the irrelevant multiplicative factor, which
only depends on $r$.

\subsection{$u\rightarrow(2+r)\eta$ limit}

We show that in the limit $u\rightarrow(2+r)\eta$, the corner transfer
matrices $B$ and $D$ are diagonal. For unshifted weights of the
RSOS model in the $u\rightarrow(2+r)\eta$ limit we get 
\begin{eqnarray}
\alpha_{l}(u=(2+r)\eta) & = & -\frac{h(\eta r)}{h(2\eta)},\label{eq:u2plusr1}\\
\beta_{l}(u=(2+r)\eta) & = & -\frac{h(2\eta-\eta r)}{h(2\eta)}\frac{[h(2\eta(l-1))h(2\eta(l+1))]^{1/2}}{h(2\eta l)},\nonumber \\
\gamma_{l}(u=(2+r)\eta) & = & -\frac{h(2\eta(l+1)-\eta r)}{h(2\eta l)},\nonumber \\
\delta_{l}(u=(2+r)\eta) & = & \frac{h(2\eta(l-1)-\eta r)}{h(2\eta l)}.\nonumber 
\end{eqnarray}
 The shifted weights take the form 
\begin{eqnarray}
\alpha_{l}^{+}(u=(2+r)\eta) & = & \alpha_{l}^{-}(u=(2+r)\eta)=0,\label{eq:u2plusr2}\\
\beta_{l}^{+}(u=(2+r)\eta) & = & -\beta_{l}^{-}(u=(2+r)\eta)=-\frac{[h(2\eta(l-1))h(2\eta(l+1))]^{1/2}}{h(2\eta l)},\nonumber \\
\gamma_{l}^{+}(u=(2+r)\eta) & = & -\gamma_{l}^{-}(u=(2+r)\eta)=-\frac{h(2\eta(l+1))}{h(2\eta l)},\nonumber \\
\delta_{l}^{+}(u=(2+r)\eta) & = & -\delta_{l}^{-}(u=(2+r)\eta)=-\frac{h(2\eta(l-1))}{h(2\eta l)}.\nonumber 
\end{eqnarray}
 Because $\alpha_{l}^{+}(u=(2+r)\eta)=\alpha_{l}^{-}(u=(2+r)\eta)=0$,
most of the NW-SE asymmetric weights are zero. The remaining NW-SE
asymmetric weights can be shown to be zero, by making use of the properties
Eqs.~\eqref{eq:u2plusr1}, \eqref{eq:u2plusr2}. This shows that
for $u\rightarrow(2+r)\eta$, the corner transfer matrices $B$ and
$D$ are diagonal.

As was the case for $u=0$, the plaquettes which contribute in the
limit $u\rightarrow(2+r)\eta$ fall in two classes, the ones with
one term and those with two terms. Again, the plaquette weights of
these two classes can be shown to give rise to the same weights, by
making use of the elliptic function relation in Eq.~\eqref{eq:elliptic-relation}.
The final form of the weights is slightly more complicated than in
the case $u=0$, and does in fact depend on to the heights, 
\begin{equation}
\tilde{W}(l_{1},l_{2},l_{3,}l_{4},l_{5},l_{6},l_{7},l_{8})(u=(2+r)\eta)=-\frac{[h(2\eta l_{3})h(2\eta l_{7})]^{1/2}}{h(2\eta l_{1})}\left(\frac{h(2\eta-\eta r)}{h(2\eta)}\right)^{2}\delta_{l_{1},l_{5}}\delta_{l_{2},l_{4}}\delta_{l_{6},l_{8}}.
\end{equation}
 Using the definition of $V_{j}$, Eq. (\ref{eq:defV}), we can show
that 
\begin{equation}
V_{j}(u=(2+r)\eta)=-\left(\frac{h(2\eta-\eta r)}{h(2\eta)}\right)^{2}\tilde{V}{}_{2j-1}\tilde{V}{}_{2j+3}\tilde{V}^{-2}{}_{2j+1}\ ,
\end{equation}
 where we introduced 
\begin{equation}
(\tilde{V}{}_{j})_{\mathbf{l},\mathbf{l}'}=[h(2\eta l_{j})]^{1/2}\delta(\mathbf{l},\mathbf{l}').
\end{equation}
 From the above and the definition of the corner transfer matrices,
Eq. (\ref{eq:cmtA}), it follows that 
\begin{equation}
B(u=(2+r)\eta)=D(u=(2+r)\eta)=\tilde{V}{}_{1},
\end{equation}
 where we again dropped the irrelevant ($l_{1}$ independent) multiplicative
factor.

\subsection{$u\rightarrow(2-r)\eta$ limit}

In this limit all the weights of the RSOS model just change sign compared
to the $u\rightarrow(2+r)\eta$ limit. Hence, the weights of the composite
model are unchanged and we find

\begin{equation}
B(u=(2-r)\eta)=D(u=(2-r)\eta)=\tilde{V}{}_{1}.
\end{equation}

\section{Weights of the composite model in the $p\rightarrow1$ limit\label{weightprop}}

\label{app:p-one-limit}

To derive the integer function $N(\mathbf{l})$, we consider the limit
$p\rightarrow1$ where weights of the composite model, and hence the
corner transfer matrix $A$, become diagonal. For $0<p<1$, employing
the conjugate modulus transformation, the function $h(u)$ can be
written in the following way \cite{abf84}:

\begin{equation}
h(u)=\tau\exp\left[-\frac{\pi(u-K)^{2}}{2KK'}\right]E(e^{-2\pi u/K'},y),
\end{equation}
 where 
\begin{align}
y & =e^{-4\pi K/K'} & \tau & =\frac{K}{K'}\prod_{n=1}^{\infty}\frac{1-y^{n/2}}{1+y^{n/2}}
\end{align}
 and the function $E(z,x)$ is Jacobi's triple product 
\begin{equation}
E(z,x)=\prod_{n=1}^{\infty}(1-x^{n-1}z)(1-x^{n}z^{-1})(1-x^{n})\ .
\end{equation}
 Then the weights of the RSOS model have the form 
\begin{align}
\alpha_{l} & =\nu\frac{g_{l}^{2}}{g_{l-1}g_{l+1}}w^{1/2}\frac{E(xw^{-1},x^{r})}{E(x,x^{r})}\nonumber \\
\beta_{l} & =\nu\frac{g_{l-1}g_{l+1}}{g_{l}^{2}}\left(\frac{xE(x^{l-1},x^{r})E(x^{l+1},x^{r})}{wE^{2}(x^{l},x^{r})}\right)^{1/2}\frac{E(w,x^{r})}{E(x,x^{r})}\label{eq:cmt-weights}\\
\gamma_{l} & =\nu\left(\frac{g_{l+1}}{g_{l}}\right)^{2}\frac{E(x^{l}w,x^{r})}{E(x^{l},x^{r})}\nonumber \\
\delta_{l} & =\nu\left(\frac{g_{l-1}}{g_{l}}\right)^{2}\frac{E(x^{l}w^{-1},x^{r})}{E(x^{l},x^{r})}\nonumber 
\end{align}
 where 
\begin{align}
x & =e^{-4\pi\eta/K'} & w & =e^{-2\pi u/K'} & g_{l}=\exp[-\frac{\pi u(2\eta l-K)^{2}}{8\eta KK'}]
\end{align}
 and $\nu$ is a constant independent of $l$.

In what follows, we show that in the limit of $p\rightarrow1$ and
$w\rightarrow1$ the weights of the composite model become diagonal
(similar to the RSOS model). It is straightforward to see that the
limit $p\rightarrow1$ implies that $x\rightarrow0$, since $K'(p\rightarrow1)$
diverges. To find out the limiting values of the weights, we use the
following properties of the function $E(z,x)$:

\begin{equation}
\lim_{x\rightarrow0}E(x^{l}w,x^{r})=\begin{cases}
1-w & l=0\\
1 & 1\leq l\leq r-1
\end{cases}
\end{equation}

\begin{equation}
\lim_{x\rightarrow0}E(wx^{l+r/2},x^{r})=\begin{cases}
1 & 0\leq l<r/2\\
1-w^{-1} & l=r/2\\
-wx^{r/2-l} & r/2<l\leq r-1
\end{cases}
\end{equation}

\begin{equation}
\lim_{x\rightarrow0}E(wx^{l-r/2},x^{r})=\begin{cases}
-wx^{l-r/2} & 0\leq l<r/2\\
1-w & l=r/2\\
1 & r/2<l\leq r-1
\end{cases}
\end{equation}

In this limit the weights of the RSOS model (unshifted as well as
shifted) can be written as

\begin{align}
\alpha_{l} & =w^{\frac{r-1}{2r}} & \alpha_{l}^{+} & =-w^{-\frac{r+1}{2r}}x^{\frac{3-r}{4}} & \alpha_{l}^{-} & =w^{\frac{r-1}{2r}}x^{\frac{1-r}{4}}\nonumber \\
\beta_{l} & =(1-w)w^{-\frac{r-1}{2r}}x^{\frac{1}{2}} & \beta_{l}^{+} & =w^{\frac{1-r}{2r}}x^{\frac{3-r}{4}} & \beta_{l}^{-} & =-w^{\frac{1+r}{2r}}x^{\frac{1-r}{4}}\label{eq:weightsp1}\\
\gamma_{l} & =w^{\frac{1+2l-r}{2r}} & \gamma_{l}^{+} & =\begin{cases}
w^{\frac{1+2l-r}{2r}}x^{\frac{1+2l-r}{4}} & l<\frac{r}{2}\\
w^{\frac{1}{2r}}(1-\frac{1}{w})x^{\frac{1}{4}} & l=\frac{r}{2}\\
-w^{\frac{1+2l-3r}{2r}}x^{\frac{1-2l+r}{4}} & l>\frac{r}{2}
\end{cases} & \gamma_{l}^{-} & =\begin{cases}
-w^{\frac{1+2l+r}{2r}}x^{\frac{-1+2l-r}{4}} & l<\frac{r}{2}\\
w^{\frac{1}{2r}}(1-w)x^{-\frac{1}{4}} & l=\frac{r}{2}\\
w^{\frac{1+2l-r}{2r}}x^{\frac{-1-2l+r}{4}} & l>\frac{r}{2}
\end{cases},\nonumber \\
\delta_{l} & =w^{\frac{1-2l+r}{2r}} & \delta_{l}^{+} & =\begin{cases}
-w^{\frac{1-2l-r}{2r}}x^{\frac{1+2l-r}{4}} & l<\frac{r}{2}\\
w^{\frac{1}{2r}}(1-\frac{1}{w})x^{\frac{1}{4}} & l=\frac{r}{2}\\
w^{\frac{1-2l+r}{2r}}x^{\frac{1-2l+r}{4}} & l>\frac{r}{2}
\end{cases} & \delta_{l}^{-} & =\begin{cases}
w^{\frac{1-2l+r}{2r}}x^{\frac{-1+2l-r}{4}} & l<\frac{r}{2}\\
w^{\frac{1}{2r}}(1-w)x^{-\frac{1}{4}} & l=\frac{r}{2}\\
-w^{\frac{1-2l+3r}{2r}}x^{\frac{-1-2l+r}{4}} & l>\frac{r}{2}
\end{cases}\nonumber 
\end{align}

Using the limits above and taking into account the exact forms for
the composite weights (see Appendix \ref{sec:plaquettes}), we are
able to show that the diagonal weights become much larger than the
off-diagonal ones. In particular, the leading $x$ behavior of the
weights of the diagonal plaquettes (those which are NE-SW symmetric)
is given by $x^{1-r/2}$, in the limit $x\rightarrow0$. We explicitly
checked that all the off-diagonal plaquettes have weights with a leading
exponent of $x$ strictly larger than $1-r/2$, showing that $A$
is diagonal when $p\rightarrow1$. So in this limit we can drop all
the off-diagonal weights and only diagonal weights contribute to the
corner transfer matrix: 
\begin{equation}
\tilde{W}(l_{1},l_{2},l_{3,}l_{4},l_{5},l_{6},l_{7},l_{8})=\frac{g_{l_{3}}g_{l_{7}}}{g_{l_{1}}g_{l_{5}}}x^{1-r/2}w^{|l_{1}-l_{5}|/4+\delta_{l_{1},l_{7}}\delta_{l_{6},l_{8}}\delta_{l_{1},l_{5}}}\delta_{l_{2},l_{8}}\delta_{l_{3},l_{7}}\delta_{l_{4},l_{6}}\ .
\end{equation}
 The dependence on $w$ of the weights in the limit $p\rightarrow1$
follows from the form of the weights $\alpha_{l}$, \emph{etc}., given
in Eq.~\eqref{eq:weightsp1}, combined with the form of the weights
given in Eq.~\eqref{eq:cmt-weights}. Note that the exponent of $w$
is always integer or half-integer.

Inserting the above in the definition $A$ in Eq.~(\ref{eq:cmtA}),
we get that 
\begin{equation}
A_{\mathbf{l},\mathbf{l}'}=\prod_{j=1}^{(m+1)/2}[\tilde{W}(l_{2j-1},l_{2j},l_{2j+1,}l_{2j+2},l_{2j},l_{2j+1},l_{2j+2},l_{2j+3})]^{j}\delta(\mathbf{l},\mathbf{l}')
\end{equation}
 and 
\begin{equation}
A_{\mathbf{l},\mathbf{l}'}=g_{l_{1}}^{-1}w^{\phi(\mathbf{l})}\delta(\mathbf{l},\mathbf{l}'),
\end{equation}
 where we have dropped the irrelevant multiplicative factor $x^{1-r/2}$
and $\phi(\mathbf{l})$ is given in Eq.~\eqref{eq:phi-1}.

\section{Connection with conformal field theory for $r>5$}

\label{app:cft-connection}

In the case $r>5$, we did not yet obtain explicit expressions for
the functions $X_{m}(q)$ with $m$ finite. However, the functions
$X_{m}(q)$, with $m$ odd, satisfy the following recursion relations
($m$ is odd by definition) 
\begin{equation}
\begin{split}X_{m}(a;b,c,d,e;q)= & q^{\frac{(m+1)(|b-1-e|/4+\delta_{b-1,c}\delta_{c,e}\delta_{b,d})}{2}}X_{m-2}(a;b-2,b-1,b,c;q)+\\
 & q^{\frac{(m+1)(|b-1-e|/4+\delta_{b-1,c}\delta_{c,e}\delta_{b,d})}{2}}X_{m-2}(a;b,b-1,b,c;q)+\\
 & q^{\frac{(m+1)(|b+1-e|/4+\delta_{b+1,c}\delta_{c,e}\delta_{b,d})}{2}}X_{m-2}(a;b,b+1,b,c;q)+\\
 & q^{\frac{(m+1)(|b+1-e|/4+\delta_{b+1,c}\delta_{c,e}\delta_{b,d})}{2}}X_{m-2}(a;b+2,b+1,b,c;q)\ ,\label{eq:Xrecursion}
\end{split}
\end{equation}
 where we define $X_{m}(a;b,c,d,e;q)$ to be zero if any of the $a,b,c,d,e$
lies outside of the range $1,2,\ldots,r-1$. In addition, $X_{m}(a;b,c,d,e;q)=0$
if $|b-c|\neq1$, $|c-d|\neq1$, $|d-e|\neq1$. Finally, $X_{1}(a;b,c,d,e;q)=q^{|a-e|/4+\delta_{a,c}\delta_{b,d}\delta_{c,e}}$,
if also $|a-b|=1$ and zero otherwise.

Using these recursion relations, one can obtain high order expansions
for $X_{m}(q)$, which allows one to identify the conformal field
theory, by direct comparison to the CFT characters. The following
exact results were obtained in this way, and hence not proven.

\subsection{The case $u>0$ (domain ${\cal D}_{1}$)}

For $u>0$, we could identify the functions $X_{m}(q)$, where $(a;b,c,d,e)$
correspond to ground state patterns, as characters of the coset model
$\frac{su(2)_{1}\times su(2)_{1}\times su(2)_{k-2}}{su(2)_{k}}$,
where $r-2=k\geq3$. The ground states for $u>0$ were discussed in
section \ref{sec:gs-u-pos}. The fields in this model are labeled
by $\Phi_{s_{2}}^{t',t,s_{1}}$ where $t'$ and $t$ correspond to
the factors $su(2)_{1}$, $s_{1}$ corresponds to $su(2)_{k-2}$ and
$s_{2}$ to $su(2)_{k}$. Because of the constraint $t'+t+s_{1}+s_{2}=0\bmod2$,
we can set $t'=1$ (both $t'$ and $t$ can take the values $1$ and
$2$). Finally, $s_{1}=1,2,\ldots,k-1$ and $s_{2}=1,2,\ldots,k+1$.
Note that we use the height values to label the fields.

To make the connection between the labels $(t',t,s_{1},s_{2})$, we
will assume that $m=4p+3$, with $p$ an integer (the case $m=4p+1$
is very similar). Because of the relation $X_{m}(a;b,c,d,e;q)=X_{m}(r-a;r-b,r-c,r-d,r-e;q)$,
we only have to consider two cases for the labels $(a;b,c,d,e)$.
The function $\lim_{p\rightarrow\infty}X_{4p+3}(a;b,b-1,b,b+1;q)$
gives the character of the field $\Phi_{a}^{1,1,b-1}$. Finally, the
function $\lim_{p\rightarrow\infty}X_{4p+3}(a;b,b+1,b+2,b+1;q)$ gives
the character of the fields $\Phi_{a}^{1,2,b}$.

For completeness, we give the scaling dimensions of the fields in
the coset theory explicitly. This formula resembles the formula for
the scaling dimensions of the minimal models \cite{bpz84}.

Finding the scaling dimensions of the fields in coset conformal field
theories is typically easiest done in a Coulomb gas formalism \cite{df84}.
In the case at hand, the relevant Coulomb gas was studied in Ref.~\cite{ijs10}.
Based on those results we find (by appropriately constraining the
values of the electric and magnetic charges) 
\begin{equation}
h(t,s_{1},s_{2})=\begin{cases}
\frac{(s_{1}(k+2)-s_{2}k)^{2}-4}{8k(k+2)}+\frac{1}{2}-\frac{(s_{1}-s_{2}+2t)\bmod4}{4} & \text{for \ensuremath{s_{1}+s_{2}\bmod2=0}}\\
\frac{(s_{1}(k+2)-s_{2}k)^{2}-4}{8k(k+2)}+\frac{1}{8} & \text{for \ensuremath{s_{1}+s_{2}\bmod2=1}}
\end{cases}
\end{equation}
 The scaling dimensions satisfy $h(3-t,k-s_{1},k+2-s_{2})=h(t,s_{1},s_{2})$,
reflexing the fact that the fields $\Phi_{k+2-s_{2}}^{3-t',3-t,k-s_{1}}$
and $\Phi_{s_{2}}^{t',t,s_{1}}$ are identified.

\subsection{The case $u<0$ (domain ${\cal D}_{2}$)}

For $u<0$, the system is described in terms of $Z_{k}$ parafermions,
and we find the following identification. The $Z_{k}$ parafermion
fields are labeled by two integers, $\Phi_{n}^{j}$, where $j=0,1,\ldots,k$,
and $j+n=0\bmod2$. Two fields which only differ in their $n$ label
by $2k$ are identified, $\Phi_{n}^{j}\equiv\Phi_{n+2k}^{j}$. In
addition, one has the identification $\Phi_{n}^{j}\equiv\Phi_{n+k}^{k-j}$,
which is reflected in the function $X_{m}(q)$ via $X_{m}(a;b,c,d,e;q)=X_{m}(r-a;r-b,r-c,r-d,r-e;q)$.
We remind the reader that $k=r-2$, and that the parameters $a,\ldots,e$
lie in the range $1,2,\ldots r-1$.

For $u<0$, the ground states are specified uniquely by $(a;b,c)$,
because $d=b$ and $e=c$ (the ground states for $u<0$ were discussed
in Section \ref{sec:gs-u-neg}). Note that $c=b\pm1$. The label $a$
corresponds directly to the label $j=a-1$. The magnitude of the field
label $n$ is given by $c-1$, with a positive sign if $c=b+1$, and
a negative sign if $c=b-1$. Thus, in general we find that $\lim_{m\rightarrow\infty}q^{(m+1)(m+3)/8}X_{m}(a;b,c,b,c;q^{-1})$
corresponds to the character of the field $\Phi_{(c-b)(c-1)}^{a-1}$,
where the limit is taken over the odd integers.


\begin{thebibliography}{10}
\bibitem{mr91} G.~Moore, N.~Read, \textit{Nonabelions in the fractional
quantum Hall effect}, Nucl. Phys. B \textbf{360}, 362 (1991).

\bibitem{ftl07} A.~Feiguin, S.~Trebst, A.W.W.~Ludwig, M.~Troyer,
A.~Kitaev, Z.~Wang, M.H.~Freedman, \textit{Interacting anyons in
topological quantum liquids: The golden chain}, Phys. Rev. Lett. \textbf{98},
160409 (2007).

\bibitem{taf08} S. Trebst, E. Ardonne, A. Feiguin, D.A. Huse, A.W.W.
Ludwig, M. Troyer, \emph{Collective States of Interacting Fibonacci
Anyons}, Phys. Rev. Lett. \textbf{101}, 050401 (2008).

\bibitem{gat09} C.~Gils, E.~Ardonne, S.~Trebst, A.W.W.~Ludwig,
M.~Troyer, Z.~Wang, \textit{Collective States of Interacting Anyons,
Edge States, and the Nucleation of Topological Liquids}, Phys. Rev.
Lett. \textbf{103}, 070401 (2009).

\bibitem{lpt11} A.W.W.~Ludwig, D.~Poilblanc, S.~Trebst, M.~Troyer,
\textit{Two-dimensional quantum liquids from interacting non-Abelian
anyons} New J. Phys. \textbf{13} 045014 (2011).

\bibitem{gs09} E.~Grosfeld, K.~Schoutens, \textit{Non-abelian anyons:
when Ising meets Fibonacci}, Phys. Rev. Lett. \textbf{103}, 076803
(2009).

\bibitem{bsh09} F.A.~Bais, J.K.~Slingerland, S.M.~Haaker, \textit{A
theory of topological edges and domain walls}, Phys. Rev. Lett. \textbf{102},
220403 (2009).

\bibitem{abf84} G.E.~Andrews, R.J.~Baxter, P.J.~Forrester, \textit{Eight-Vertex
SOS model and generalized Rogers-Ramanujan-type identities}, J. Stat.
Phys. \textbf{35}, 193 (1984).

\bibitem{book:b82} R.J.~Baxter, \textit{Exactly solved models in
statistical mechanics}, Academic Press, London (1982).

\bibitem{h84} D.A.~Huse, \textit{Exact exponents for infinitely
many new multicritical points}, Phys. Rev. B \textbf{30}, 3908 (1984).

\bibitem{sb89} H.~Saleur, M.~Bauer, \emph{On some relations between
local height probabilities and conformal invariance}, Nucl. Phys.
B \textbf{320}, 591 (1989).

\bibitem{djk87} E.~Date, M.~Jimbo, A.~Kuniba, T.~Miwa, M.~Okado,
\textit{Exactly solvable SOS models. Local height probabilities and
theta function identities}, Nucl. Phys. B \textbf{290} 231 (1987).

\bibitem{kkm93} R.~Kedem, R.R.~Klassen, B.M.~McCoy, E.~Melzer,
\textit{Fermionic sum representations for conformal field theory characters},
Phys. Lett. B \textbf{307}, 68 (1993).

\bibitem{h91} F.D.M.~Haldane, \textit{``Fractional statistics''
in arbitrary dimensions: a generalization of the Pauli principle},
Phys. Rev. Lett. \textbf{67}, 937 (1991).

\bibitem{s97} K.~Schoutens, \textit{Exclusion Statistics in Conformal
Field Theory Spectra}, Phys. Rev. Lett. \textbf{79}, 2608 (1997).

\bibitem{gs99} S.~Guruswamy, K.~Schoutens, \textit{Non-abelian
Exclusion Statistics}, Nucl. Phys. B \textbf{556}, 530 (1999).

\bibitem{ijs09} Y.~Ikhlef, J.L.~Jacobsen, H.~Saleur, \textit{A
Temperley-Lieb quantum chain with two- and three-site interactions},
J. Phys. A \textbf{42}, 292002 (2009).

\bibitem{ijs10} Y.~Ikhlef, J.L.~Jacobsen, H.~Saleur, \textit{The
$Z_{2}$ staggered vertex model and its applications}, J. Phys. A
\textbf{43}, 225201 (2010).

\bibitem{longsu2k} C.~Gils, E.~Ardonne, S.~Trebst, D.A.~Huse,
A.W.W.~Ludwig, M.~Troyer, Z.~Wang, in preparation.

\bibitem{ttw08} S.~Trebst, M.~Troyer, Z.~Wang, A.W.W.~Ludwig,
\textit{A short introduction to Fibonacci anyon models}, Prog. Theor.
Phys. Supp. \textbf{176}, 384 (2008).

\bibitem{kr88} A.N.~Kirillov, N.Y.~Reshetikhin, \textit{Representations
of the algebra $U_{q}(sl(2))$, $q$-orthogonal polynomials and invariants
of links}, in V.G.~Kac, ed., \textit{Infinite dimensional Lie algebras
and groups, Proceedings of the conference held at CIRM, Luminy, Marseille},
p. 285, World Scientific, Singapore (1988).

\bibitem{mg69a} C.K.~Majumdar, D.K.~Ghosh, \textit{On Next-Nearest-Neighbor
interaction in linear chain. I}, J. Math. Phys. \textbf{10}, 1388
(1969).

\bibitem{mg69b} C.K.~Majumdar, D.K.~Ghosh, \textit{On Next-Nearest-Neighbor
interaction in linear chain. II}, J. Math. Phys. \textbf{10}, 1399
(1969).

\bibitem{longj3} S.~Trebst, D.A.~Huse, E.~Ardonne, A.~Feiguin,
A.W.W.~Ludwig, M.~Troyer, in preparation.

\bibitem{g87} D.~Gepner, \textit{New conformal field theories associated
with lie algebras and their partition functions,} Nucl. Phys. B \textbf{290},
10 (1987).

\bibitem{zf85} A.B.~Zamolodchikov, V.A.~Fateev V A, \textit{Nonlocal
(parafermion) currents in two-dimensional conformal quantum Þeld theory
and self-dual critical points in $Z_{N}$-symmetric statistical systems},
Zh. Eksp. Teor. Fiz. \textbf{89}, 380 (1985).

\bibitem{p87} V. Pasquier, \textit{Two-dimensional critical systems
labelled by Dynkin diagrams}, Nucl. Phys. B \textbf{285}, 162 (1987).

\bibitem{book:kbi93} V.E.~Korepin, N.M.~Bogoliubov, A.G.~Izergin,
\emph{Quantum Inverse Scattering Method and Correlation Functions},
Cambridge University Press, Cambridge (1993).

\bibitem{ks93} W.M.~Koo, H.~Saleur, \textit{Fused Potts models},
Int. Jour. Mod. Phys. A \textbf{8}, 5165 (1993).

\bibitem{j83} V.F.R.~Jones, \textit{Index for subfactors}, Invent.
Math. \textbf{72}, 1 (1983).

\bibitem{w87} H.~Wenzl, \textit{On sequences of projections}, C.R.
Math. Rep. Acad. Sci. Canada \textbf{9}, 5 (1987).

\bibitem{b07} R.J.~Baxter, \textit{Corner transfer matrices in statistical
mechanics}, J. Phys. A \textbf{40}, 12577 (2007).

\bibitem{arr01} E.~Ardonne, N.~Read, E.~Rezayi, K.~Schoutens,
\textit{Non-Abelian quantum Hall states: wave functions and quasihole
state counting}, Nucl. Phys. B \textbf{607}, 549 (2001).

\bibitem{as99} E.~Ardonne, K.~Schoutens, \textit{New class of non-abelian
spin-singlet quantum Hall states}, Phys. Rev. Lett. \textbf{82}, 5096
(1999).

\bibitem{g95up} G.~Georgiev, \textit{Combinatorial constructions
of modules for infinite-dimensional Lie algebras, II. Parafermionic
space}, arXiv:q-alg/9504024, (1995).

\bibitem{js06} J.L.~Jacobsen, H.~Saleur, \textit{The antiferromagnetic
transition for the square-lattice Potts model}, Nucl. Phys. B \textbf{743},
207 (2006).

\bibitem{agl11} E.~Ardonne, J.~Gukelberger, A.W.W.~Ludwig, S.~Trebst,
M.~Troyer, \textit{Microscopic models of interacting Yang-Lee anyons,}
New J. Phys. \textbf{13}, 045006 (2011).

\bibitem{fb85} P.J.~Forrester, R.J.~Baxter, \textit{Further exact
solutions of the eight-vertex SOS model and generalizations of the
Rogers-Ramanujan indentities}, J. Stat. Phys. \textbf{38}, 435 (1985).

\bibitem{bpz84} A.A.~Belavin, A.M.~Polyakov, A.B.~Zamolodchikov,
\textit{Infinite conformal symmetry in two-dimensional quantum field
theory}, Nucl. Phys. B \textbf{241}, 333 (1984).

\bibitem{df84} Vl.S.~Dotsenko, V.A.~Fateev, \textit{Conformal algebra
and multipoint correlation functions in 2D statistical models}, Nucl.
Phys. B \textbf{240}, 312 (1984).\end{thebibliography}
\end{document}